\documentclass[aps,preprint,12pt,nofootinbib]{revtex4-1}

\usepackage{lipsum}
\usepackage{amsmath}
\usepackage{amssymb}
\usepackage{physics}
\usepackage{setspace}
\usepackage{latexsym, color,dsfont}
\usepackage{comment}

\usepackage{xcolor}
\usepackage{graphicx}
\usepackage{dcolumn}
\usepackage{caption}
\usepackage{bm}
\usepackage{subcaption}
\usepackage{mwe}
\usepackage[normalem]{ulem}
\usepackage{hyperref}

\newcommand{\be}{\begin{eqnarray}}
\newcommand{\ee}{\end{eqnarray}}
\begin{document}

\title{Black Hole Information From Non-vacuum Localised Quantum States}

\author{Ali Akil$^{1,2}$, Riccardo Falcone$^{3*}$,  Nicetu {Tibau Vidal}$^{1,2}$, \\
Giulio Chiribella$^{1,2}$}
\affiliation{
$^1$QICI Quantum Information and Computation Initiative, Department of Computer Science, The University of Hong Kong, Pok Fu Lam Road, Hong Kong\\
$^2$HKU-Oxford Joint Laboratory for Quantum Information and Computation\\
$^3$Department of Physics, University of Sapienza, Piazzale Aldo Moro 5, 00185 Rome, Italy}
\email{ aliakil@hku.hk,
riccardo.falcone@uniroma1.it, nicetu@hku.hk, giulio@hku.hk }

\begin{abstract}
We revisit Hawking's black hole radiation derivation, including the quantum state of the initial matter forming the black hole. We investigate how non-vacuum initial quantum states, at the past of a black hole geometry, influence the 
black hole radiation observed at future null infinity $( \mathcal{I}^+)$. We further classify which of the initial state excitations are distinguishable from one another through measurements on the black hole radiation state.
We use Algebraic Quantum Field Theory (AQFT) to provide a clear physical interpretation of the results, in terms of localised operations. 
We then take a concrete example of a black hole made of one large collapsing excitation of mass $M$ and compare it to a same-mass black hole formed due to the collapse of two smaller excitations, of mass $M/2$ each. We find using our formalism that the two cases yield different radiation states and can in principle be distinguished.
Our results provide a mechanism for partial information recovery in evaporating black holes, classify what information is recoverable through stimulated emission, and a concrete understanding of the classification based on the AQFT localisation. 
\end{abstract}

\maketitle

\section{Introduction}
In 1975, Stephen Hawking showed that black holes emit radiation. Comparing the vacuum state at the future null infinity $ \mathcal{I}^+$ with that at the past null infinity $\mathcal{I}^-$, one can see that the past infinity's vacuum contains particles according to the future infinity's definition of particles. The radiation was shown to be in a two-mode squeezed state, one mode inside the black hole and another one outside. 
As the black hole produces these pairs, its mass decreases, and the horizon shrinks. If nothing stops this `evaporation', the black hole will eventually cease to exist and turn completely into radiation \cite{Hawking74,Parker:1975jm, BirrelDavies, Fabbri}. 
The main problem arises from the fact that the state of the Hawking radiation seems to be independent of the detailed state of the matter that formed the black hole and only carries information about the total black hole mass (as well as total angular momentum and total electric charge when they exist). Moreover, although the Hawking radiation is in a pure state when one includes the future horizon modes, once the black hole fully evaporates, the horizon modes no longer exist to purify the radiation state. This leaves us with only the exterior (`out') radiation, which is in a mixed state \cite{Hawking76,HawkingInfo}. This appears to hold even when the original matter that formed the black hole is in a pure state. This pure-to-mixed-state evolution in a closed system violates unitarity, one of the fundamental principles of Quantum Theory. 
Attempts to solve this information loss problem are countless. They include black hole remnants \cite{Adler:2001vs,Stable,Perez,Rovelli2}, complementarity \cite{Susskind1,Susskind2,Complementarity}, final-state projection \cite{Maldacena, Lloyd, Harlow}, fuzzballs \cite{Mathur}, 
firewalls \cite{AMPS,Braunstein},  islands \cite{Penington}, wormholes \cite{Almheiri:2019qdq,Penington:2019kki}, entanglement swapping \cite{Akil1} and many others. For an extensive review on quantum information in black holes the reader could refer to \cite{BOOK}.

Here, we will take a bottom-up approach in the spirit of stimulated black hole emission. 
Stimulated emission in rotating black holes was originally introduced in \cite{Press:1972zz,Starobinsky:1973aij}, where it was shown that bosons entering and exiting a Kerr black hole's ergosphere induce the creation of more bosons. Then, after Hawking studied particle production in Schwarzschild black hole spacetime backgrounds, it was shown that the Hawking radiation can be enhanced by considering the infalling matter's quantum state \cite{Wald:1976ka}. 
The probability distribution and energy spectrum of the stimulated radiation emission were calculated in \cite{Bekenstein:1977mv} and \cite{Panangaden:1977pc}. It was there understood that the stimulated-emitted particles' temperature is shifted towards the temperature of the infalling radiation that stimulated it. This point, in particular, was more emphasized in Sorkin's \cite{Sorkin:1986zj}, where a simple derivation of stimulated emission was provided. However, this applies only to the thermal infalling radiation and not a general state. The localisation of infalling matter was discussed in \cite{Audretsch:1994ga}, thus making the discussion slightly more rigorous. The algebraic quantum field theory localisation, however, was not yet well known and widely understood; the authors hence implicitly used a Modal localisation scheme, damping each wave mode outside its desired support.
Using the Brandenberger-Mukhanov-Prokopec method \cite{Brandenberger:1992sr} for coarse-grained entropy computations, \cite{Keski-Vakkuri:1993hsn} derived a counterintuitive result showing that although stimulated emission increases the number of emitted particles, their entropy decreases. The (1+1)-dimensional black hole case was discussed in \cite{Vendrell:1996am} with a focus on BTZ black holes. Lochan, Padmanabhan, and Chakraborty
discussed what information could be extracted from the radiation in the stimulated emission case; they found specific infalling states with particular symmetry constraints, for which the information can be fully recovered from the outgoing radiation \cite{Lochan:2015oba, Lochan:2016nbs}.
Our computations here may provide a clear interpretation of their work. Up to that point, however, stimulated emission had been treated mainly as a thermodynamic effect. The first instance we know of, using the language of quantum states and quantum channels, is in \cite{Bradler:2013gqa}. Based on arguments from information capacity of quantum channels, they showed that classical information can be preserved due to stimulated emission. 

In this article, we will build upon all of the above. First, we will derive the quantum states of the outgoing radiation, first formally, for any initial/infalling quantum state in scalar quantum field theory. Then, we will derive the concrete expressions of the Hawking radiation quantum states in the cases of the single particle and the coherent initial quantum state. We will classify what type of states can possibly be recovered in the Hawking radiation and what others seem to be lost. We find that the states differing from each other through a unitary operator from the algebra of the Horizon part of the future Cauchy hypersurface, will not be distinguishable. On the other hand, those that differ from each other through a non-unitary from the same algebra, or any operator (regardless unitary or not) from the null infinity algebra, will at least in principle be distinguishable. For the sake of a precise interpretation of the result, we discuss the localisation of ingoing states invoking a rigorous localisation scheme from algebraic quantum field theory. 
We find that the fact that the initial states differing from each other (even at $ \mathcal{I}^-$) by an operator from the `out' algebra are distinguishable is a trivial statement. The reason is that these `out' operations represent operations that do not at any point fall into the black hole, even when considering such `out' operators acting on the `in' vacuum state at $ \mathcal{I}^-$. The fact that an operation from the black hole horizon's (`int') algebra, if not a local unitary, can still in principle be recovered from the outside radiation, turns out to be due to the Knight-Licht and the impossibility of localizing a non-unitarily prepared state. We consider more elaborately, in the light of these findings, the case of single particle excitations in Vaidya spacetime, showing that those with different energies cannot be related to each other through local unitaries. This implies that they are distinguishable through measurements on the black hole radiation. 
Then we consider the infalling matter forming the black hole with mass M, in two different configurations. The first is two particles with mass M/2 each, and the second is a single particle with mass M. We derive the radiation states that the two cases yield and show that they are different `out' states, thus preserving information. 
Although this does not recover the full fledged unitarity, that seems to be lost during black hole evaporation, yet
it clearly demonstrates how stimulated emission allows for the recovery of information in the radiation state up to a unitary acting in the infalling region.

We will establish the general framework in \ref{General_case}, look at the Vaidya spacetime example in \ref{Vaidya_spacetime}, and discuss general state distinguishability  in \ref{sec:disting}. Section \ref{sec:Non-vacuum} derives the form of a general ``in'' state, in the `out-int' basis and classifies what operations on the ``in'' states leave an imprint on the Hawking radiations and which operations do not (\ref{Vacuum_vs_excitation}), then discusses the distinguishability of initial states through radiation measurements (\ref{excitations_vs_excitations}).  Section \ref{subsec:nosignal} establishes the physical interpretation of our results. Section \ref{localisation} focuses on the Knight-Licht localisability of unitary operations while \ref{AQFT}
connects the Fock space bipartition to localisation through algebraic quantum field theory. In \ref{Single}, we compare infalling single particles with different energies. Then in \ref{twoparticles} we compare infalling matter of the same energy but different numbers of particles. Finally, conclusions are laid out in \ref{sec:conclusions}.

\section{Framework}
In this section, we establish the framework upon which we will build our results. We will start with the most general possible treatment, then focus on the special example of a Vaidya spacetime. Our results should be valid for any other spacetime with a dynamic (time-dependent) region. 
\subsection{General framework}\label{General_case}

This paper considers the asymptotically flat spacetime of a black hole in formation. We denote the past null infinity, future null infinity, and the horizon by $\mathcal{I}^-$, $ \mathcal{I}^+$ and $\mathcal{H}^+$, respectively. Additionally, we consider a massless scalar field $\hat{\phi}$ that satisfies the Klein-Gordon equation $\square \hat{\phi} = 0$ and the commutation relation
\begin{equation}\label{Phi_commutation}
\left[ \hat{\phi}(x), \hat{\phi}(x') \right] = i \hbar \Delta(x-x'),
\end{equation}
where $x = (x^0, \vec{x})$ represents any spacetime coordinate system and $\Delta(x)$ is defined as the retarded minus the advanced propagator. Explicitly, $\Delta(x)$ satisfies the Klein-Gordon equation $ \square \Delta(x) = 0$, with the initial conditions $ \sqrt{- g} g^{0 \mu} \left. \partial_\mu \Delta \right|_{x^0=0} = \delta^3(\vec{x})$ and $ \left. \Delta \right|_{x^0=0} = 0 $, where $g_{\mu\nu}$ is the spacetime metric and $g$ is its determinant.

The field $\hat{\phi}$ may be decomposed into a basis of Klein-Gordon modes that are orthonormal with respect to the Klein-Gordon inner product 

\begin{equation} \label{KleinGordon_product}
(f,f') = i \int_\Sigma d\Sigma \sqrt{h} n^\mu\!\!\!_{_\Sigma} \left( f^* \partial_\mu f' - f' \partial_\mu f^* \right),
\end{equation}
with $\Sigma$ as any Cauchy hypersurface, $n^\mu_\Sigma$ as the unit normal to $\Sigma$ and $h$ as the determinant of the induced spatial metric. By the equations of motion, the definition (\ref{KleinGordon_product}) does not depend on the specific choice for the Cauchy surface \cite{Jacobson:2003vx}.

We consider two distinct decompositions of the quantum field:
\begin{subequations}
\begin{align}
& \hat{\phi} = \int_0^{+\infty} d\omega \left[ \left( \hat{\phi}, u^\text{in}_\omega \right) u^\text{in}_\omega + \text{h.c.} \right], \\
& \hat{\phi} = \int_0^{+\infty} d\omega \left[ \left( \hat{\phi}, u^\text{out}_\omega \right) u^\text{out}_\omega + \left( \hat{\phi}, u^\text{int}_\omega \right) u^\text{int}_\omega + \text{h.c.} \right].\label{phi_out_int}
\end{align}
\end{subequations}
Here, $u^\text{in}_\omega$ are the ``in'' modes, defined as the orthonormal solutions of the Klein-Gordon equation with positive frequency on $\mathcal{I}^-$. Conversely, the ``out'' modes $u^\text{out}_\omega$ are defined with positive frequency on $\mathcal{I}^+$. On the surface $\mathcal{I}^+ \cup \mathcal{H}^+$, the $u^\text{out}_\omega$ modes are assumed to be supported exclusively on $\mathcal{I}^+$ and do not extend to $\mathcal{H}^+$. To account for the portion of the surface that is not covered by $u^\text{out}_\omega$, additional modes $u^\text{int}_\omega$ are introduced. These modes are supported on $\mathcal{H}^+$ and vanish on $\mathcal{I}^+$. On $\mathcal{I}^-$, the regions $\mathcal{I}^-_\text{out} \subset \mathcal{I}^-$ and $\mathcal{I}^-_\text{int} \subset \mathcal{I}^-$ are defined as the respective supports of $u^\text{out}_\omega$ and $u^\text{int}_\omega$ for all $\omega$. This implies that $u^\text{out}_\omega$ vanishes on $\mathcal{I}^-_\text{int}$ and $u^\text{int}_\omega$ vanishes on $\mathcal{I}^-_\text{out}$ for any $\omega$.

In addition to $\hat{\phi}$, we define its canonical conjugate $\hat{\pi}_\Sigma$ with respect to any Cauchy surface $\Sigma$ as $\hat{\pi}_\Sigma = \sqrt{h} n^\mu_\Sigma \partial_\mu \hat{\phi}|_{\Sigma} $, where $n^\mu_\Sigma$ is the unit normal to $\Sigma$ and $h$ is the determinant of the induced spatial metric on $\Sigma$. The decompositions of $\hat{\pi}_\Sigma$ with respect to the ``in'', ``out'' and ``int'' modes are
\begin{subequations}
\begin{align}
& \hat{\pi}_\Sigma = \int_0^{+\infty} d\omega \left[ \left( \hat{\phi}, u^\text{in}_\omega \right) \sqrt{h} n^\mu_\Sigma \partial_\mu u^\text{in}_\omega|_{\Sigma} + \text{h.c.} \right], \\
& \hat{\pi}_\Sigma = \int_0^{+\infty} d\omega \left[ \left( \hat{\phi}, u^\text{out}_\omega \right) \sqrt{h} n^\mu_\Sigma \partial_\mu u^\text{out}_\omega|_{\Sigma} + \right. \nonumber\\ & \quad\left.+\left( \hat{\phi}, u^\text{int}_\omega \right) \sqrt{h} n^\mu_\Sigma \partial_\mu u^\text{int}_\omega|_{\Sigma} + \text{h.c.} \right].\label{pi_out_int}
\end{align}
\end{subequations}

The operators $\hat{a}^\text{in}_\omega = \left( \hat{\phi}, u^\text{in}_\omega \right)$, $\hat{a}^\text{out}_\omega = \left( \hat{\phi}, u^\text{out}_\omega \right)$ and $\hat{a}^\text{int}_\omega = \left( \hat{\phi}, u^\text{int}_\omega \right)$ are interpreted as the annihilators for particles associated with the modes $u^\text{in}_\omega$, $u^\text{out}_\omega$ and $u^\text{int}_\omega$ respectively \footnote{Within the formalism of AQFT, the annihilators $\hat{a}^{\text{in}}_\omega$, $\hat{a}^{\text{out}}_\omega$ and $\hat{a}^{\text{int}}_\omega$ are operators acting on the corresponding Fock space $\mathcal{F}_\text{in}$, $\mathcal{F}_\text{out}$ and $ \mathcal{F}_\text{int}$. They are related by the fact that the quantum operators $ \int _{0}^{+\infty} d \omega ( u^{\text{in}}_\omega \hat{a}^{\text{in}}_\omega + \text{h.c.} ) $ and $ \int_0^{+\infty} d\omega (  u^{\text{out}}_\omega \hat{a}^{\text{out}}_\omega +  u^\text{int}_\omega \hat{a}^{\text{int}}_\omega + \text{h.c.})$ offer two different representations of the same element of the field algebra \cite{haag1992local}.
}
. These operators are related via the Bogoliubov transformations
\begin{gather}\label{Bogoliubov}
\hat{a}^\text{in}_\omega = \int_0^{+\infty} d\omega' \left[ \left(  u^\text{out}_{\omega'} , u^\text{in}_\omega \right)  \hat{a}^\text{out}_{\omega'} + \left(  u^\text{int}_{\omega'} , u^\text{in}_\omega \right) \hat{a}^\text{int}_{\omega'} + \right. \nonumber \\ \left.+ \left(  u^{\text{out} *}_{\omega'} , u^\text{in}_\omega \right)  \hat{a}^{\text{out}  \dagger}_{\omega'} + \left(  u^{\text{int}*}_{\omega'} , u^\text{in}_\omega \right) \hat{a}^{\text{int} \dagger}_{\omega'}  \right].
\end{gather} 
The respective vacuum states $| \text{in} \rangle$, $ | \text{out} \rangle$, and $ | \text{int} \rangle$ are defined by the conditions $\hat{a}^\text{in}_\omega | \text{in} \rangle = 0$, $ \hat{a}^\text{out}_\omega | \text{out} \rangle = 0$ and $\hat{a}^\text{int}_\omega | \text{int} \rangle = 0$ for all $\omega$. We assume the existence of an operator $\hat{S}$ such that \footnote{In AQFT, the vectors $| \text{in} \rangle$ and $\hat{S} | \text{out} \rangle \otimes | \text{int} \rangle$ are two distinct representation of the same state, each belonging to a different Hilbert space: $\mathcal{F}_\text{in}$ and $\mathcal{F}_\text{out} \otimes \mathcal{F}_\text{int}$, respectively \cite{haag1992local}. }
\begin{equation}\label{in_S_out_int}
| \text{in} \rangle = \hat{S} | \text{out} \rangle \otimes | \text{int} \rangle.
\end{equation}
This operator can be derived using the condition  $\hat{a}^\text{in}_\omega | \text{in} \rangle = 0$ for all $\omega$, along with the Bogoliubov transformation (\ref{Bogoliubov}). The resulting Hawking state is obtained by taking the partial trace of $| \text{in} \rangle \langle \text{in} |$ over the ``int'' degrees of freedom:
\begin{equation}\label{rho_H}
\hat{\rho}_\text{H} = \text{Tr}_\text{int} \left[  \hat{S} | \text{out} \rangle  \langle \text{out} | \otimes | \text{int} \rangle  \langle \text{int} | \hat{S} \right].
\end{equation}

\subsection{A particular example: Vaidya spacetime}\label{Vaidya_spacetime}

The results presented in this paper are applicable to the general scenario described in Sec.\ \ref{General_case}. However, it is often useful to examine specific cases where all physical quantities are explicitly defined. As a concrete example that satisfies the assumptions outlined in Sec.\ \ref{General_case}, we consider the Vaidya spacetime, representing a Schwarzschild black hole formed by the collapse of incoming radiation. This spacetime is defined by the metric
\begin{equation}
ds^2 = - \left[1- \frac{2 M \theta(v - v_0)}{r} \right] dv^2 + 2 dv dr + r^2 d\Omega^2
\end{equation}
where $v=t+r$ is the ingoing light-cone coordinate, $r$ is the radial coordinate, $d\Omega^2$ represents the standard Riemannian metric of the unit 2-sphere, $M$ is the mass of the black hole and $v_0$ denotes the $v$-coordinate corresponding to incoming radiation. The Minkowski region $\mathcal{M}$ is identified by $v<v_0$, whereas the Schwarzschild region $\mathcal{S}$ by $v>v_0$. In $\mathcal{M}$, we consider the null coordinates $(u_\text{in}, v)$, with $u_\text{in} = v - 2 r$. In $\mathcal{S}$, instead, it is convenient to use $(u_\text{out}, v)$, with $u_\text{out} = v - 2 r^*$ and with $r^*$ defined by $dr^*=(1-2M/r)^{-1}dr$. 
The past null infinity $\mathcal{I}^-$ is identified by $u_\text{in} \to -\infty$ and $u_\text{out} \to -\infty$ in the respective regions $\mathcal{M}$ and $\mathcal{S}$; the future null infinity $\mathcal{I}^+$ is given by the limit $v \to \infty$; the horizon $\mathcal{H}^+$ is identified by $u_\text{in} = v_0 - 4M$ and  $u_\text{out} \to \infty$.

The ``in'' modes $u^\text{in}_\omega$, as defined in Sec.\ \ref{General_case}, are the orthonormal solutions to the Klein-Gordon equation that satisfy the boundary condition $ u^\text{in}_\omega|_{\mathcal{I}^-} = e^{- i \omega v}/4 \pi \sqrt{\omega} r$. Additionally, they vanish in $\mathcal{M}$ as $r \to 0$. In the Minkowski region $\mathcal{M}$, their explicit form is
\begin{equation}\label{u_in_M}
\left. u^\text{in}_\omega \right|_\mathcal{M} = \frac{1}{4 \pi \sqrt{\omega} r} \left[ e^{-i \omega v} - e^{-i \omega u_\text{in}} \right].
\end{equation}
In the Schwarzschild region $\mathcal{S}$, under near-horizon approximation, the modes $u^\text{in}_\omega$  are given by 
\begin{equation}\label{u_in_S}
\left. u^\text{in}_\omega \right|_\mathcal{S} = \frac{1}{4 \pi \sqrt{\omega} r} \left[ e^{-i \omega v} - e^{-i \omega u_\text{in}(u_\text{out},v_0)} \right].
\end{equation}
This expression solves the Klein-Gordon equation $\partial_{u_\text{out}} \partial_v (r u^\text{in}_\omega) = 0$ and satisfies the boundary conditions $u^\text{in}_\omega = [e^{-i \omega v_0} - e^{-i \omega u_\text{in}(u_\text{out},v_0)} ]/4 \pi \sqrt{\omega} r$ at $v = v_0$ and $u^\text{in}_\omega \to e^{-i \omega v} / 4 \pi \sqrt{\omega} r$ as $u_\text{out} \to -\infty$ and $v>v_0$. Here, $u_\text{in}(u_\text{out},v_0)$ is the inverse of  $u_\text{out}(u_\text{in},v_0) = u_\text{in} - 4 M \log | (v_0-u_\text{in})/2 - 2 M |$.

While the $u^\text{in}_\omega$ modes have a positive frequency on $\mathcal{I}^-$, the ``out'' modes $u^\text{out}_\omega$ are defined with a positive frequency on $\mathcal{I}^+$.  In the respective regions, these modes are expressed as
\begin{align}\label{u_out}
& \left. u^\text{out}_\omega \right|_\mathcal{S} = \frac{e^{- i \omega u_\text{out}}}{4 \pi \sqrt{\omega} r}, \nonumber \\ & \left. u^\text{out}_\omega \right|_\mathcal{M} =\frac{1}{4 \pi \sqrt{\omega} r} \left[ e^{- i \omega u_\text{out}(u_\text{in},v_0)} - \theta(v_H - v) e^{- i \omega u_\text{out}(v,v_0)} \right],
\end{align}
where $v_H = v_0 - 4 M $ represents the $v$-coordinate of the null ray forming the event horizon at $u_\text{out} = + \infty$ \cite{Fabbri}. On the other hand, the ``int'' modes are given by
\begin{equation}\label{u_int}
u^\text{int}_\omega = - \frac{\theta(v-v_H)}{4 \pi \sqrt{\omega} r} \exp \left\lbrace i \omega \left[ v_H - 4 M \log \left( \frac{v-v_H}{4 M} \right) \right] \right\rbrace.
\end{equation}
When restricted to $\mathcal{I}^-$, the ``out'' and the ``int'' modes are supported in the regions separated by $v = v_H$. Specifically, $\mathcal{I}^-_\text{out}$ corresponds to $v < v_H$, while $\mathcal{I}^-_\text{int}$ corresponds to $v > v_H$.

\begin{figure}[ht]
    \centering
    \includegraphics[width=0.5\linewidth]{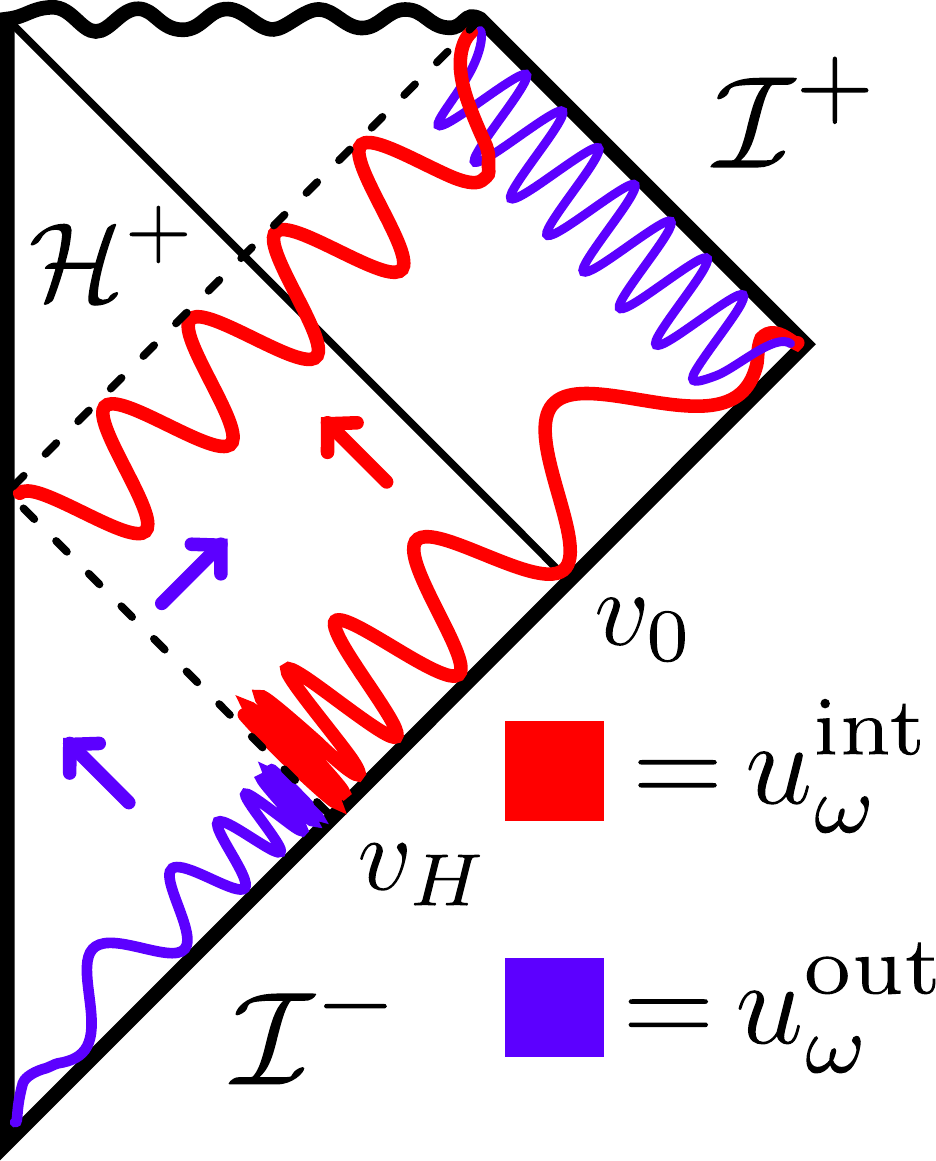}
    \caption{A Penrose diagram of the Schwarzschild black hole. $\mathcal I^-$ is the past null-like infinity, $\mathcal I^+$ is the future null-like infinity, the dashed line is the null line that forms the event horizon $\mathcal H^+$, and the wiggly line is the black hole singularity. The figure distinguishes wave modes (in red) that at the past infinity satisfy the conditions that make them ``int'' modes in the future, and the wave modes (in blue) that make them ``out'' modes in the future. Among the possible operations on the initial state (at $\mathcal I^-$), only a unitary on the modes in red will not directly impact the radiation. A non-unitary on the modes in red, as well as any operation on the modes in blue will leave an imprint on the radiation.   }
    \label{fig:penrose}
\end{figure}

By using the form of the modes provided in Eqs.\ (\ref{u_in_M}), (\ref{u_in_S}), (\ref{u_out}) and (\ref{u_int}), the Bogoliubov coefficients in Eq.\ (\ref{Bogoliubov}) can be explicitly computed. The detailed steps of these calculations are omitted here, as they are not essential to the discussion and can be found in standard references, such as \cite{Fabbri}. The crucial result is that these Bogoliubov transformations lead to Eq.\ (\ref{in_S_out_int}), where the operator $\hat{S}$ takes the form of a two-mode squeezing operator. The resulting state satisfies the identities (see Eqs.\ (3.152) and (3.153) in Ref.\ \cite{Fabbri})
\begin{align}
    & \left( \hat{a}^\text{out}_\omega - e^{-4\pi M \omega} \hat{a}^{\text{int}\dagger}_\omega \right) | \text{in} \rangle = 0, \label{a_out_a_int_dagger_0_in} \\
    & \left( \hat{a}^\text{int}_\omega - e^{-4\pi M \omega} \hat{a}^{\text{out}\dagger}_\omega \right) | \text{in} \rangle = 0,
\end{align}
Taking the partial trace of $| \text{in} \rangle \langle \text{in} | $ over the ``int'' Fock space produces a thermal density matrix with temperature $T_\text{H} = 1/8\pi k_\text{B} M$, resulting in a thermalized Hawking state 

\be \label{eq:hawking}
\hspace{-4mm} \rho_H =\bigotimes_\omega  \left(1-e^{-8 \pi \omega M}\right) \! \! \sum_{n_\omega=0}^{+\infty} \! \! e^{- 8\pi n_\omega \omega M} \ketbra{n_\omega}^{\rm out}. 
\ee
\subsection{Distinguishability and Information Recovery} \label{sec:disting}

The information paradox arises from the belief that the radiation state in Eq.(\ref{eq:hawking}) only depends on the total black hole mass, apparently implying that regardless what the fine-grained initial black hole matter state is, the evaporation seems to erase that information and yield the same final state. 
In other words, initial states, based on the common interpretation, seem not to be distinguishable from measurements on the final state.

First of all we need to remove the ambiguity by clarifying what we mean by distinguishability here. In fact, quantum states are said to be only `distinguishable' when they are orthogonal. However, what that is referring to is the one-shot measurement distinguishability, which is only possible in the case where the full information encoded in the initial state is fully recoverable from the radiation state. Although we do not recover this one-shot measurement distinguishability in our work, we still rigorously recover a significant part of the information encoded in the initial state.
We will henceforth call any two systems {\it tomographically distinguishable} if their density matrices are different. More formally, we call two out states 
$\rho$ and $\sigma$ tomographically distinguishable if and only if $\rho \neq \sigma$. 

Note that Hawking's many-to-one evolution, if true, would imply that even with infinite resources and infinite copies of the black hole and its radiation, one still cannot reconstruct initial states, because their corresponding radiation (final) states are not tomographically distinguishable. This makes the theory not conserve information besides the total mass. This asymptotic or tomographic distinguishability is what we are recovering here up to an `int' unitary, as we will show.
We will find that two initial states $\ket{\psi_{\text{in}}}$ and $\ket{\varphi_{\text{in}}}$ will lead to two tomographically distinguishable final states if there is no unitary ``int'' operator $\hat{U}_\text{int}$ such that $\ket{\psi_{\text{in}}} = \hat{U}_\text{int} \ket{\varphi_{\text{in}}}$.

\section{Radiation from Non-vacuum initial states} \label{sec:Non-vacuum}
In this section, we consider excitations of the ``in'' vacuum and we address the following questions: What states can be completely absorbed by the black hole without leaving any trace in the radiation? And what ``in'' vacuum excitations can be recovered from the radiation? We find that the ``in'' region quantum states that are generated by applying a unitary transformation to the vacuum using a unitary operator from the algebra of ``int'' creators and annihilators leave no trace in the Hawking radiation. We then ask what excited (non-vacuum) initial states are distinguishable from one another. The answer similarly turns out to be that ``int''-unitarily equivalent initial states are not distinguishable from each other, everything else is.

\subsection{Vacuum vs excitation}\label{Vacuum_vs_excitation}

Any excitation of the ``in'' vacuum can be expressed as $| \psi \rangle = \hat{O}  | \text{in} \rangle$, where  $\hat{O} = O_\text{in}(\hat{a}^\text{in}_\omega, \hat{a}^{\text{in} \dagger}_\omega) $ is an element of the algebra generated by $\hat{a}^\text{in}_\omega$ and $\hat{a}^{\text{in}^\dagger}_\omega$. Using the Bogoliubov transformation (\ref{Bogoliubov}), the operator $O_\text{in}(\hat{a}^\text{in}_\omega, \hat{a}^{\text{in} \dagger}_\omega)$ can be rewritten in terms of the ``out'' and ``int'' annihilation and creation operators, since these form a complete basis.  This allows us to define $O_\text{out,int}(\hat{a}^\text{out}_\omega, \hat{a}^{\text{out} \dagger}_\omega,\hat{a}^\text{int}_\omega, \hat{a}^{\text{int} \dagger}_\omega)$ such that $O_\text{in}(\hat{a}^\text{in}_\omega, \hat{a}^{\text{in} \dagger}_\omega) = O_\text{out,int}(\hat{a}^\text{out}_\omega, \hat{a}^{\text{out} \dagger}_\omega,\hat{a}^\text{int}_\omega, \hat{a}^{\text{int} \dagger}_\omega)$. Combining this with Eq.\ (\ref{in_S_out_int}), the state $| \psi \rangle$ can be expressed as an element of the ``out'' and ``int'' Fock space $\mathcal{F}_{\text{out},\text{int}}$:
\begin{equation}\label{psi_O_out_int}
| \psi \rangle = O_\text{out,int} \left( \hat{a}^\text{out}_\omega, \hat{a}^{\text{out} \dagger}_\omega,\hat{a}^\text{int}_\omega, \hat{a}^{\text{int} \dagger}_\omega \right) \hat{S} | \text{out} \rangle \otimes | \text{int} \rangle.
\end{equation}

We now investigate the conditions under which the state $| \psi \rangle$ is indistinguishable from the vacuum $| \text{in} \rangle$ by measurements in the ``out'' region, such that it does not perturb the Hawking radiation. We use the Schrödinger–HJW theorem \cite{Schrodinger1936,HUGHSTON199314}, which states that for any couple of ``in'' states, their partial traces are equal, i.e., $\Tr_{\text{int}}\left(\ketbra{\psi_{\text{in}}}\right)=\Tr_{\text{int}}\left(\ketbra{\varphi_{\text{in}}}\right)$, iff there exists a unitary ``int'' operator $U_{\text{int}}$ such that $\ket{\psi_{\text{in}}}= \left(U_{\text{int}} \otimes \mathbb{I}_{\text{out}}\right) \ket{\varphi_{\text{in}}}$. In our case, we identify $\ket{\psi_{\text{in}}}= \ket{\psi}$ and $\ket{\varphi_{\text{in}}}= \ket{\text{in}}$, so that $\text{Tr}_\text{int}  | \psi \rangle  \langle \psi | = \hat{\rho}_\text{H}$ iff there exists a unitary operator $\hat{U}_\text{int}$ such that $O_\text{out,int} ( \hat{a}^\text{out}_\omega, \hat{a}^{\text{out} \dagger}_\omega,\hat{a}^\text{int}_\omega, \hat{a}^{\text{int} \dagger}_\omega) = \mathbb{I}_\text{out} \otimes \hat{U}_\text{int}$. This condition is satisfied when the perturbing operator $\hat{O}$ is unitary and composed solely of ``int'' annihilation and creation operators. As a result, we find that the states that leave the Hawking state unperturbed are $| \psi \rangle =  \mathbb{I}_\text{out} \otimes \hat{U}_\text{int} | \text{in} \rangle$, with $\hat{U}_\text{int} \in \mathfrak{A}_\text{int}$ as a unitary element of the algebra $\mathfrak{A}_\text{int}$ generated by $\hat{a}^\text{int}_\omega$ and $\hat{a}^{\text{int}^\dagger}_\omega$.

The opposite situation arises when the operator $\hat{O}$ consists solely of ``out'' annihilation and creation operators. In this case, $O_\text{out,int} ( \hat{a}^\text{out}_\omega, \hat{a}^{\text{out} \dagger}_\omega,\hat{a}^\text{int}_\omega, \hat{a}^{\text{int} \dagger}_\omega )$ can be written as $O_\text{out,int} ( \hat{a}^\text{out}_\omega, \hat{a}^{\text{out} \dagger}_\omega,\hat{a}^\text{int}_\omega, \hat{a}^{\text{int} \dagger}_\omega ) = \hat{O}_\text{out} \otimes \mathbb{I}$, where $ \hat{O}_\text{out} \in \mathfrak{A}_\text{out}$. By plugging this expression into
\be
\label{Tr_int_psi}
\text{Tr}_\text{int}  | \psi \rangle  \langle \psi | && = \text{Tr}_\text{int} \left\lbrace O_\text{out,int} \left( \hat{a}^\text{out}_\omega, \hat{a}^{\text{out} \dagger}_\omega,\hat{a}^\text{int}_\omega, \hat{a}^{\text{int} \dagger}_\omega \right) \right. \nonumber \\
&& \left. \hat{S} | \text{out} \rangle \langle \text{out} | \otimes | \text{int} \rangle \langle \text{int} | \hat{S}^\dagger \left[ O_\text{out,int} \left( \hat{a}^\text{out}_\omega, \hat{a}^{\text{out} \dagger}_\omega,\hat{a}^\text{int}_\omega, \hat{a}^{\text{int} \dagger}_\omega \right) \right]^\dagger \right\rbrace.
\ee
We find that
\begin{equation}\label{Tr_psi_O_out}
\text{Tr}_\text{int}  | \psi \rangle  \langle \psi | = \hat{O}_\text{out} \hat{\rho}_H \hat{O}_\text{out}^\dagger.
\end{equation}
In this scenario, the operator $\hat{O}_\text{out}$ is completely unaffected by the partial trace and appears in $\mathcal{I}^+$ as a perturbation acting on the Hawking radiation, regardless whether this operator is unitary or not. 

As an example of an excitation of the ``in'' vacuum, consider a single ``in'' particle state, defined by
\begin{equation}
O_\text{in} (\hat{a}^\text{in}_\omega, \hat{a}^{\text{in} \dagger}_\omega) = \int_0^{+\infty} d\omega \psi_\omega \hat{a}^{\text{in} \dagger}_\omega,
\end{equation}
where $\psi_\omega$ represents the particle's wave function. In this case, the operator $\hat{O}$ is nonunitary, implying that the particle inevitably perturbs the Hawking state, regardless of the specific form of $\psi_\omega$.

As an example of a unitary excitation of the ``in'' vacuum,  consider coherent states, defined by
\begin{equation}
O_\text{in} (\hat{a}^\text{in}_\omega, \hat{a}^{\text{in} \dagger}_\omega) = \exp ( \int_0^{+\infty} d\omega \psi_\omega \hat{a}^{\text{in} \dagger}_\omega - \text{h.c.}).
\end{equation}
In this case, the operator $\hat{O}$ is unitary. Hence, the condition for the state $| \psi \rangle$  to be indistinguishable from the ``in'' vacuum $| \text{in} \rangle$ in the ``out'' region is that $\hat{O}$ must consist entirely of ``int'' annihilation and creation operators.

\subsection{Excitation vs excitation}
\label{excitations_vs_excitations}In Sec.\ref{Vacuum_vs_excitation}, we introduced an excited ``in'' state $| \psi \rangle = \hat{O}  | \text{in} \rangle$ and examined conditions under which its resulting ``out'' radiation differs from the Hawking state $\hat{\rho}_\text{H}$. 

The conditions under which the two partial traces $\text{Tr}_\text{int}  | \psi \rangle  \langle \psi |$ and $\text{Tr}_\text{int}  | \psi' \rangle  \langle \psi' |$ are equivalent, are given again by the Schrödinger–HJW theorem \cite{Schrodinger1936, HUGHSTON199314}.  The theorem states that $\text{Tr}_\text{int}  | \psi \rangle  \langle \psi | = \text{Tr}_\text{int}  | \psi' \rangle  \langle \psi' |$ if and only if there is a unitary $\hat{U}_\text{int} = U_{\rm int}( \hat{a}^{\rm int}_\omega, \hat{a}^{\rm int \dagger}_\omega )$ in the ``int'' algebra that relates one to another, i.e.,
\be \label{eq:ExcExc}
|\psi \rangle = U_{\rm int} ( \hat{a}^{\rm int}_\omega, \hat{a}^{\rm int \dagger}_\omega ) |\psi' \rangle.
\ee

We find that the two states are indistinguishable in the ``out'' region, only if Eq.\ \eqref{eq:ExcExc} holds. In every other case, the two excitations are distinguishable in principle from measurements in the ``out'' region. Namely, 
for 
\be \label{NonUniInt}
|\psi \rangle = O_\text{int}( \hat{a}^{\rm int}_\omega, \hat{a}^{\rm int \dagger}_\omega ) |\psi' \rangle,
\ee
where $\hat{O}_\text{int}  = O_\text{int}( \hat{a}^{\rm int}_\omega, \hat{a}^{\rm int \dagger}_\omega )$ is non-unitary, the imprints of the two states on the ``out'' radiation are distinguishable, as they yield a different marginal state on the ``out'' Hilbert space.
Similarly, but perhaps less physically surprising (for reasons that will be clarified with localisation), excitations that differ from each other through a regardless unitary or not ``out'' operator. That is, for 
\be \label{OutExc}
|\psi \rangle = O_\text{out}( \hat{a}^{\rm int}_\omega, \hat{a}^{\rm int \dagger}_\omega ) |\psi' \rangle,
\ee
the states $|\psi \rangle$ and $|\psi' \rangle$ are distinguishable in the ``out'' region.

From all this analysis one can deduce that initial states in general are partially reconstructable from ``out'' region final state measurements, that is, reconstructable up to an ``int'' algebra unitary transformation. This implies that a significant part of the information encoded in the black hole matter can indeed be recovered.

\section{localisation as a physical interpretation}
\label{subsec:nosignal}

\subsection{Partition, causality, and localisability } \label{localisation}
Among operations, be it unitaries or not, from the ``int'' and ``out'' algebras and the consequent bipartition of the Hilbert space $\mathcal{F}_{\text{out},\text{int}} \cong \mathcal{F}_\text{out} \otimes  \mathcal{F}_\text{int}$, we have distinguished which operations can leave a trace on the radiation and which ones do not, when acting on an ``in'' state. However, it is {\it a priori} not clear what it physically means for ``int'' and ``out'' operators to act on the ``in'' states.
In order to study the physical interpretation of the findings of Sec. \ref{sec:Non-vacuum}, we resort to the AQFT framework. We switch our focus from creation and annihilation operators ($\hat{a}^\text{int}_\omega$, $\hat{a}^\text{out}_\omega$) to field operators ($\hat{\phi}$, $\hat{\pi}$), thus allowing for a genuine localisation of operations.

The creation and annihilation operators $\hat{a}^\text{int}_\omega$ and $\hat{a}^\text{out}_\omega$ produce a bipartition of the Hilbert space $\mathcal{F}_{\text{out},\text{int}} \cong \mathcal{F}_\text{out} \otimes  \mathcal{F}_\text{int}$. Likewise, in AQFT, field operators $\phi(x)$ and $\pi(x)$ produce a factorization of the global Hilbert space into local Hilbert spaces, associated to disconnected subregions of a given Cauchy surface. In this section, we find that the bipartition $\mathcal{F}_{\text{out},\text{int}} \cong \mathcal{F}_\text{out} \otimes  \mathcal{F}_\text{int}$ matches one of these local factorizations of the global Hilbert space. Within the Cauchy surface $\mathcal{H}^+ \cup \mathcal{I}^+$, the region associated to $\mathcal{F}_\text{out}$ is $\mathcal{I}^+$, while $\mathcal{F}_\text{int}$ is associated to $\mathcal{H}^+$. Conversely, in the past infinity $\mathcal{I}^-$, the local Hilbert space $\mathcal{F}_\text{out}$ ($\mathcal{F}_\text{int}$) is associated to $\mathcal{I}^-_\text{out}$ ($\mathcal{I}^-_\text{int}$).

The regions $\mathcal{I}^-_\text{out}$ and $\mathcal{I}^-_\text{int}$ have been defined in Sec.\ \ref{General_case} as the support of the ``out'' and ``int'' modes in the infinite past. However, they also capture the causal structure of the space-time, as they represent the causal past of  $\mathcal{I}^+$ and $\mathcal{H}^+$, respectively. This a consequence of the fact that the time evolution of modes, such as those satisfying the Klein-Gordon equation, respects the light-cone structure, as in the classical theory. At least classically, any light ray that falls into the black hole cannot reach future infinity, whereas rays that evade the black hole before the horizon forms eventually appears in $\mathcal{I}^+$. Consequently, $\mathcal{I}^-$ is divided into the two subregions, $\mathcal{I}^-_\text{int}$ and $\mathcal{I}^-_\text{out}$, which identify null rays at initial times that will fall into the black hole or surpass it, respectively.

We show that the ``out'' operators represent operations acting in a region of spacetime where even classical geodesics completely avoid the black hole. This makes it more or less trivial that the ``out'' operations do leave a trace on the black hole radiation. Moreover, the fact that an ``int'' unitary does not leave a trace on the ``out'' radiation can be 
understood by using the theory Knight and Licht \cite{10.1063/1.1703731, 10.1063/1.1703925}, who define strictly localized states as those whose local preparation in a region does not affect measurements in spacelike-separated regions. The state $| \psi \rangle = \hat{O}_1 | \text{in} \rangle$ is said to be localized in $\mathcal{O}_1$ if, for any observable $\hat{O}_2$ localized in $\mathcal{O}_2$ (spacelike-separated from $\mathcal{O}_1$), we find that $\langle \psi | \hat{O}_2 | \psi \rangle = \langle \text{in} | \hat{O}_2 | \text{in} \rangle$. Due to microcausality, $\hat{O}_1$ and $\hat{O}_2$ commute (this can be seen from Eq.\ (\ref{Phi_commutation}), as $\Delta(x-x')$ vanishes when $x$ and $x'$ are spacelike separated); hence $\langle \psi | \hat{O}_2 | \psi \rangle = \langle \text{in} | \hat{O}_1^\dagger  \hat{O}_1 \hat{O}_2 | \text{in} \rangle$. However, the ``in'' vacuum $| \text{in} \rangle$ is known to be entangled across the regions $\hat{O}_1$ and $\hat{O}_2$, as indicated by the Reeh-Schlieder theorem and related results \cite{haag1992local,Ric}. Then, the hypothesis of strict localisation requires that $\hat{O}_1$ is unitary.

\subsection{AQFT localisation scheme}\label{AQFT}

Here we apply the AQFT notion of localisation to give a physical interpretation of the operators $\hat{U}_\text{int}$, $\hat{O}_\text{int}$ and $\hat{O}_\text{out}$ as localized operations on $\mathcal{I}^-_\text{int}$ and $\mathcal{I}^-_\text{out}$.
We rely on the fact that, for any Cauchy surface $\Sigma$, any quantum field operator $\hat{O}$ can be expressed as a combination of products of operators of the form of
\begin{equation}\label{O_f_Sigma}
\hat{A}[f_{\hat{\phi}\Sigma},f_{\hat{\pi}\Sigma}] = \int_{\Sigma} d^3 \vec{x} \left( f_{\hat{\phi}\Sigma} \hat{\phi} |_{\Sigma} + f_{\hat{\pi}\Sigma} \hat{\pi}_{\Sigma} \right),
\end{equation}
where $\vec{x}$ denotes spatial coordinates on $\Sigma$  and $f_{\hat{\phi}\Sigma}$ and $f_{\hat{\pi}\Sigma}$ are functions on $\Sigma$. In the AQFT localisation scheme, $\hat{A}[f_{\hat{\phi}\Sigma},f_{\hat{\pi}\Sigma}]$ is considered spatially localized on the union of the supports of $f_{\hat{\phi}\Sigma}$ and $f_{\hat{\pi}\Sigma}$.

Let us now consider the ``out'' operator $\hat{a}^\text{out}_\omega$ and the ``int'' operator $\hat{a}^\text{int}_\omega$ which have been defined in Sec.\ref{General_case} as $\hat{a}^\text{out}_\omega = \left( \hat{\phi}, u^\text{out}_\omega \right)$ and $\hat{a}^\text{int}_\omega = \left( \hat{\phi}, u^\text{int}_\omega \right)$. We compute the Klein-Gordon product [Eq.\ (\ref{KleinGordon_product})] selecting $\mathcal{I}^-$ as the Cauchy surface for this derivation and obtain
\begin{subequations}\label{a_out_a_int_phi_pi}
\begin{align}
& \hat{a}^\text{out}_\omega = \int_{\mathcal{I}^-} dx^3  \left(  \hat{\phi} |_{\mathcal{I}^-}  n^\mu_{\mathcal{I}^-} \partial_\mu u^\text{out}_\omega |_{\mathcal{I}^-}  -   u^{\text{out}*}_\omega |_{\mathcal{I}^-} \hat{\pi}_{\mathcal{I}^-} \right), \\
& \hat{a}^\text{int}_\omega = \int_{\mathcal{I}^-} dx^3  \left(   \hat{\phi} |_{\mathcal{I}^-}  n^\mu_{\mathcal{I}^-} \partial_\mu u^\text{int}_\omega |_{\mathcal{I}^-} -   u^{\text{int}*}_\omega |_{\mathcal{I}^-}  \hat{\pi}_{\mathcal{I}^-} \right). 
\end{align}
\end{subequations}
Here $\vec{x}$ denotes a coordinate system used for $\mathcal{I}^-$. For instance, in the case of Vaidya spacetime, $\vec{x} = (v,\theta,\varphi)$. 

By comparing Eq.\ (\ref{a_out_a_int_phi_pi}) with Eq.\ (\ref{O_f_Sigma}), we find that $\hat{a}^\text{int}_\omega$ is equal to the right-hand side of Eq.\ (\ref{O_f_Sigma}) when $f_{\hat{\phi}\mathcal{I}^-} = n^\mu_{\mathcal{I}^-} \partial_\mu u^\text{int}_\omega |_{\mathcal{I}^-}$ and $f_{\hat{\pi}\mathcal{I}^-} = -   u^{\text{int}*}_\omega |_{\mathcal{I}^-}$. According to AQFT, at initial times, the operator is localized in the union of the supports of $ n^\mu_{\mathcal{I}^-} \partial_\mu u^\text{int}_\omega |_{\mathcal{I}^-}$ and $-   u^{\text{int}*}_\omega |_{\mathcal{I}^-}$. That union of supports is equivalent to $\mathcal{I}^-_\text{int}$ by the definition of $\mathcal{I}^-_\text{int}$ in Sec. \ref{General_case}. Similarly, it can be shown that the operator $\hat{a}^\text{out}_\omega$ is initially localized in $\mathcal{I}^-_\text{out}$. As a result, since the ``int'' operators $\hat{U}_\text{int}$ and $\hat{O}_\text{int}$ are functions of $\hat{a}^\text{int}_\omega$ and $\hat{a}^{\text{int}^\dagger}_\omega$ operators, they represents local operations in $\mathcal{I}^-_\text{int}$. Any state that is unitarily prepared in $\mathcal{I}^-_\text{int}$ at the initial time will not perturb the final Hawking radiation. Conversely, if the operator acting on the ``in'' vacuum $| \text{in} \rangle$ is localized in $\mathcal{I}^+_\text{out}$, then, at future infinity $\mathcal{I}^+ \cup \mathcal{H}^+$, it will be localized in $\mathcal{I}^+$ and, as a result, will be entirely unaffected by the partial trace over the ``int'' region.

\section{Collapsing matter information can be recovered}
\label{Collapse}

\subsection{Single int excitations}
\label{Single}

Here we analyse the specific example of two single-particle states, infalling into the black hole, with different energies $\omega$ and $\omega'$ and show that they display different radiation states. 

We start by considering the single particle state 
\be | \psi_\omega \rangle = N_\omega \hat{a}_{\omega}^{\text{int} \dagger} \ket{ {\text{in}}}. 
\ee
$N_\omega$ is the normalisation factor.  The state $ | \psi_\omega \rangle$ has two properties: (i) it is a single ``in''-particle excitation (this can be seen by noting that $\hat{a}_{\omega}^{\text{int} \dagger}$ is linear in the ``in'' creators and annihilators) and (ii) it is a state that can be locally prepared in the $\mathcal{I}^-_\text{int}$ region by acting over the ``in'' vacuum via the local operation $\hat{a}_{\omega}^{\text{int} \dagger}$.

Using Eq.\ (\ref{a_out_a_int_dagger_0_in}), we obtain the relation 
\be | \psi_\omega \rangle = N_\omega  e^{4\pi M \omega} \hat{a}^\text{out}_\omega | \text{in} \rangle ~,
\ee 
which expresses $| \psi_\omega \rangle$ as an ``out'' operator acting on the ``in'' vacuum. Since such an operator remains unchanged under the partial trace over the ``int'' space, we find \be 
\text{Tr}_\text{int} | \psi_\omega \rangle \langle \psi_\omega | = N_\omega^2 e^{8\pi M \omega} \hat{a}^\text{out}_\omega \hat{\rho}_\text{H} \hat{a}^{\text{out}\dagger}_\omega ~. 
\ee 
For any other single particle state $| \psi_{\omega'} \rangle = N_{\omega'} \hat{a}_{\omega'}^{\text{int} \dagger} \ket{ {\text{in}}}$,  the corresponding partial trace is given by $ \text{Tr}_\text{int} | \psi_{\omega'} \rangle \langle \psi_{\omega'} | = N_{\omega'}^2 e^{8\pi M \omega'} \hat{a}^\text{out}_{\omega'} \hat{\rho}_\text{H} \hat{a}^{\text{out}\dagger}_{\omega'} $. When $\omega\neq\omega'$, these expressions differ since the Hawking state, $\rho_H$, is an uncorrelated thermal state shown in Eq.(\ref{eq:hawking}), confirming that single-particle states with different momenta lead to distinct radiation states. In what follows we elaborate on this idea.

In Sec.\ \ref{sec:Non-vacuum}, we derived the necessary and sufficient conditions for two global states to display the same outside radiation state. Applying this criterion to $| \psi_\omega \rangle$ and $| \psi_{\omega'} \rangle$, we find that the condition is not satisfied. Specifically, it can be shown that
\begin{equation}\label{a_omega_neq_a_omega_prime}
  N_{\omega} \hat{a}_{\omega}^{\text{int } \dagger} \ket{ {\text{in}}}\neq \left(U_{\text{int}} \otimes \mathbb{I}_{\text{out}} \right) N_{\omega '}\hat{a}_{\omega '}^{\text{int } \dagger} \ket{ {\text{in}}}  ~,
\end{equation}
for any $\omega \neq \omega '$, $N_{\omega}, N_{\omega'}$ normalization factors, and any unitary that is local in the interior region $U_{\text{int}}$. The detailed proof of Eq.\ (\ref{a_omega_neq_a_omega_prime}) is provided in Appendix \ref{Proof_appendix}. This condition is equivalent to the ``out'' reduced states being distinct, namely, $\rho_\omega^{\text{out}}\neq \rho_{\omega'}^{\text{out}}$.

As a result, single particle states of different energies $\omega \neq \omega'$ are not (local-)unitarily equivalent. Consequently,
 if a black hole is formed by a collapsing matter state consisting of a single scalar ``int'' mode excitation of momentum $\omega$, then different momenta will exhibit different radiation states. This result is expected since even the usual Hawking radiation from the vacuum state depends on the collapsing matter's total mass $M$.

\subsection{Different mass distributions}
\label{twoparticles}
Now we consider two states with the same total mass but different particle numbers. We show they yield different radiation in the outside region. Specifically, we consider the following two normalized states \be 
&&\ket{\psi_1} = N_M \hat{a}^{\text{int} \dagger}_{M} \ket{ {\text{in}}} \\ 
&& \ket{\psi_2}= N_{\frac{M}{2} \frac{M}{2}} \left(\hat{a}^{\text{int} \dagger}_{\frac{M}{2}} \right)^2 \ket{ {\text{in}}}.
\ee Using Eq.\ (\ref{a_out_a_int_dagger_0_in}) and the commutation property of ``int'' and ``out'' operators, we obtain 
\be N_M \hat{a}^{\text{int} \dagger}_{M} \ket{ {\text{in}}} = N_M e^{4\pi M^2} \hat{a}^{\text{out}}_{M} \ket{ {\text{in}}}~,
\ee and 
\be N_{\frac{M}{2} \frac{M}{2}} \left(\hat{a}^{\text{int} \dagger}_{\frac{M}{2}} \right)^2 \ket{ {\text{in}}} && = N_{\frac{M}{2} \frac{M}{2}} e^{2\pi M^2}  \hat{a}^{\text{int} \dagger}_{\frac{M}{2}} \hat{a}^{\text{out}}_{\frac{M}{2}} \ket{ {\text{in}}} \nonumber \\ 
&& = N_{\frac{M}{2} \frac{M}{2}} e^{2\pi M^2} \hat{a}^{\text{out}}_{\frac{M}{2}} \hat{a}^{\text{int} \dagger}_{\frac{M}{2}} \ket{ {\text{in}}} \nonumber \\ 
&& = N_{\frac{M}{2} \frac{M}{2}} e^{4 \pi M^2} \left( \hat{a}^{\text{out}}_{\frac{M}{2}} \right)^2 \ket{ {\text{in}}}~.
\ee 
The  corresponding  partial  traces for these states are  
\be \rho^{\rm out}_M = |N_M|^2 e^{8\pi M^2} \hat{a}^{\text{out}}_{M} \hat{\rho}_\text{H} \hat{a}^{\text{out}\dagger}_{M} 
\ee 
and 
\be \rho^{\rm out}_{\frac{M}{2}\frac{M}{2} } = |N_{\frac{M}{2} \frac{M}{2}}|^2 e^{8\pi M^2}  \left( \hat{a}^{\text{out}}_{\frac{M}{2}} \right)^2 \hat{\rho}_\text{H} \left( \hat{a}^{\text{out}\dagger}_{\frac{M}{2}} \right)^2 ~.
\ee 
These expressions differ due to $\rho_H$ being uncorrelated between different frequencies, thus confirming that the two initial states lead to distinct radiation states in the outside region.

\begin{figure}[ht]
    \centering
    \begin{subfigure}[b]{0.23\textwidth}
        \centering
        \includegraphics[width=\linewidth]{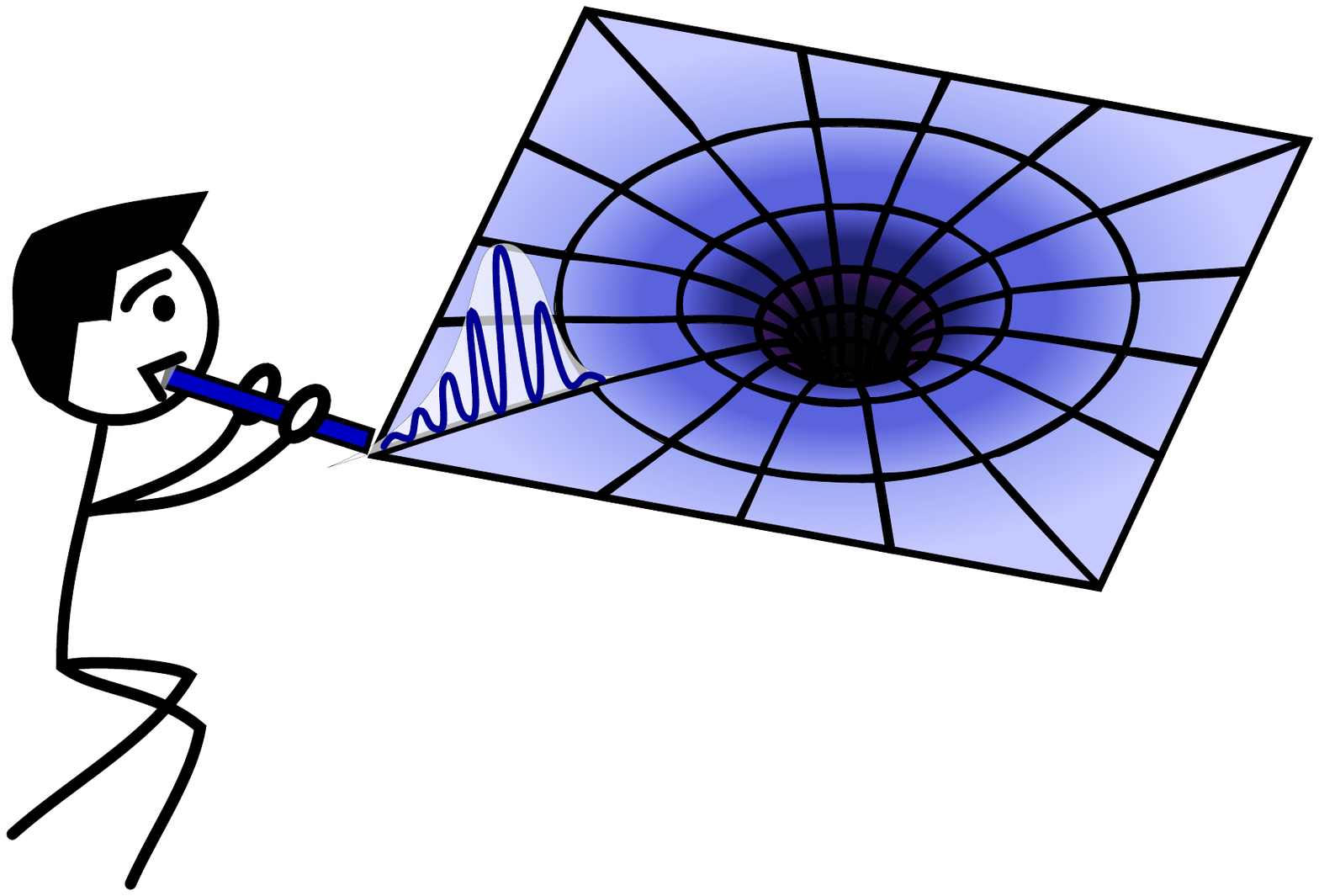}
        \label{subfig:earlypurpe}
    \end{subfigure}
    \medspace 
    \begin{subfigure}[b]{0.23\textwidth}
        \centering
        \includegraphics[width=\linewidth]{latepurple.png}
        \label{subfig:latepurpe}
    \end{subfigure}
    
    \begin{subfigure}[b]{0.23\textwidth}
        \centering
        \includegraphics[width=\linewidth]{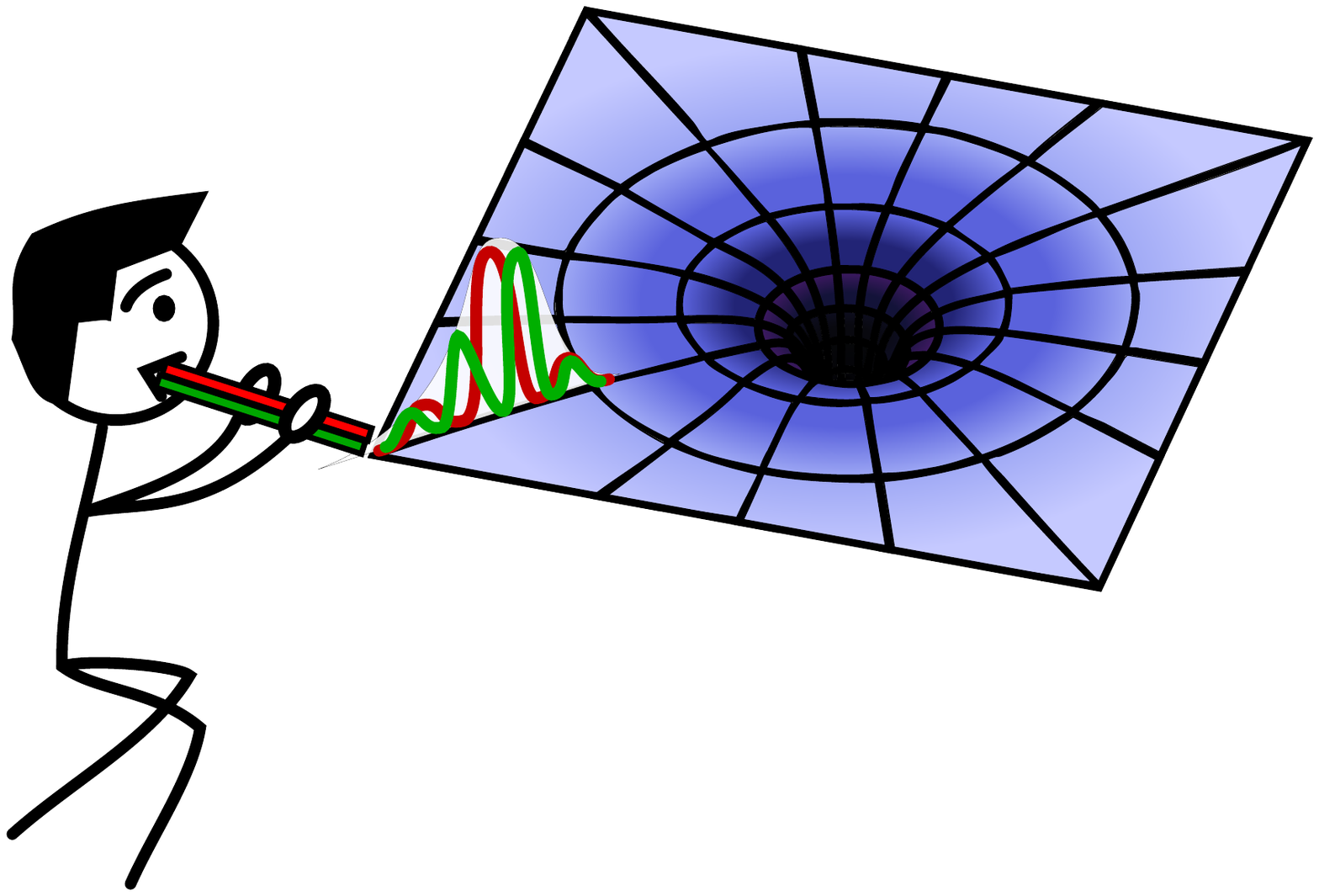}
        \label{subfig:earlygren}
    \end{subfigure}
    \medspace 
    \begin{subfigure}[b]{0.23\textwidth}
        \centering
        \includegraphics[width=\linewidth]{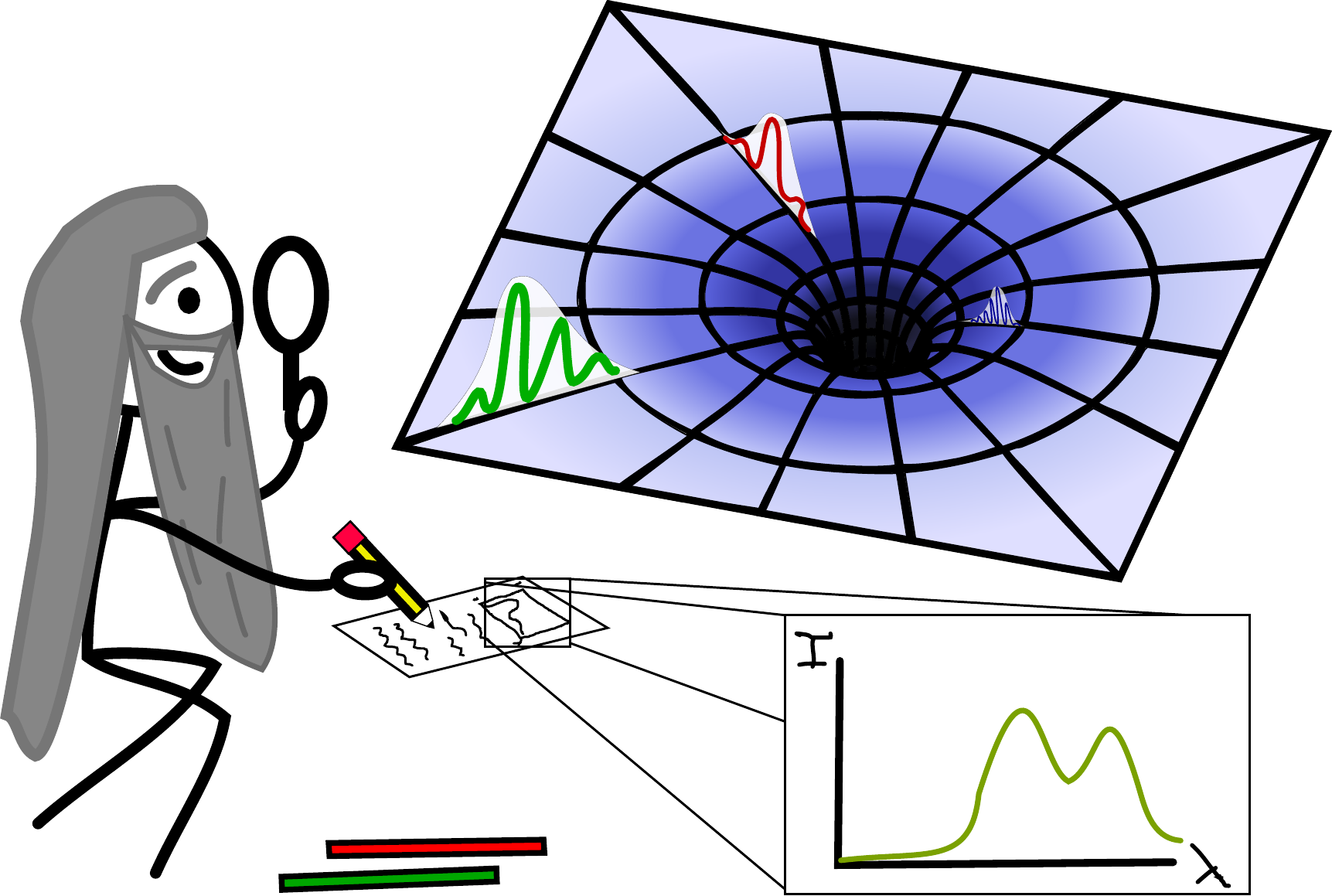}
        \label{subfig:lategreen}
    \end{subfigure}
    \caption{
    Probing black hole radiation through distinct initial-state configurations. (a) Federico sends a single matter wavepacket of mass \( M \) into the black hole. Since a one-particle ``int'' state cannot be generated by an ``int'' unitary transformation, this configuration imprints a signature on the outgoing radiation. 
    (b) After a billion generations, Ahmad detects radiation with a spectrum deviating from spontaneous Hawking emission. The radiation spectrum measured by Ahmad is observably (though not one-shot observably) different from the (spontaneously emitted) Hawking radiation. (c) 
    In another instance, Andreu sends two wavepackets into a black hole, each of mass $M/2$. Knowing that 1-particle and 2-particle ``int'' states are not equivalent through an ``int'' unitary, Andreu knows that this will leave a different imprint on the black hole radiation from the one left by Federico's wavepacket. (d) Francesco later observes radiation with a spectrum distinct from both Hawking radiation and Ahmad’s measurement, demonstrating that inequivalent ``int'' states produce observable and distinguishable modifications to the black hole’s emission over cosmological timescales.}
    \label{fig:unified_caption}
\end{figure}

An alternative way to prove the same result, analogously to Eq.\ (\ref{a_omega_neq_a_omega_prime}), is showing that the two states are not equivalent through an ``int'' unitary. 

\begin{equation}\label{a_omega_neq_a_omega_prime_2}
  N_M \hat{a}^{\text{int } \dagger}_{M} \ket{ {\text{in}}}\neq \left(U_{\text{int}} \otimes \mathbb{I}_{\text{out}} \right) N_{\frac{M}{2} \frac{M}{2}} \left(\hat{a}^{\text{int } \dagger}_{\frac{M}{2}} \right)^2 \ket{ {\text{in}}}.
\end{equation}
The details of the proof are given in Appendix \ref{Proof_appendix_2}. 

It is clear that despite both states carrying a total energy $M$ in the ``int'' modes, the state in the ``out'' region is different. Thus, if an ``out'' observer could characterize the radiation state, they could discern the two initial states of the collapsing matter. Specifically, just by looking at the expectation value of the number operator $\hat{N}_{\omega}^{\text{out}} = \hat{a}^{\text{out}\dagger}_\omega \hat{a}^{\text{out}}_\omega$, the observer would be able to see a difference. In both cases, the observer would see that $\langle \hat{N}_{\omega}^{\text{out}} \rangle=\left(e^{8\pi M \omega}-1\right)^{-1}$ for $\omega \neq\{M, \frac{M}{2}\}$, agreeing with the Hawking thermal expected value for a black hole of mass $M$. These frequencies have not been tampered with in the initial state compared to the vacuum $\ket{\text{in}}$. If the initial state is the single particle of energy $M$, the expected value of the ``out'' number operator for momentum $M$ would differ, but the modes with a frequency $\frac{M}{2}$ will remain as in the Hawking state. Conversely, if the initial state is the 2-particle state of $\frac{M}{2}$ momentum excitations,  the expected value of the ``out'' number operator for momentum $M$ would remain as in the Hawking state, but $\frac{M}{2}$ will differ. The exact values are shown in Table \ref{tab:exp}. 
\begin{table}[h]
    \centering
    \begin{tabular}{c|c c}
         & $\langle \hat{N}_{\omega}^{\text{out}} \rangle_{N_M \hat{a}^{\text{int } \dagger}_{M} \ket{ {\text{in}}}}$ & $\langle \hat{N}_{\omega}^{\text{out}} \rangle_{N_{\frac{M}{2} \frac{M}{2}} \left(\hat{a}^{\text{int } \dagger}_{\frac{M}{2}} \right)^2 \ket{ {\text{in}}}}$ \\ \hline
         $\omega=\frac{M}{2}$  & $\left(e^{4 \pi M^2 }-1\right)^{-1}$ & $3 \left(e^{4 \pi M^2 }-1\right)^{-1}$\\ $\omega=M$ & $2 \left(e^{8 \pi M^2 }-1\right)^{-1}$  &  $\left(e^{8 \pi M^2 }-1\right)^{-1}$
    \end{tabular}
    \caption{The table shows expectation values of the black hole radiation for various incident ``int'' states. It illustrates the difference between the ``out'' radiation states for the cases where incident matter states are made of a single particles vs two particles and when the particles have mass $M/2$ vs when they have masses $M$. 
    }
    \label{tab:exp}
\end{table}

This result implies that different mass distributions of the same total mass given by a scalar field that is precisely infalling into the black hole are distinguishable by measuring the ``out'' radiation. This result, as to be expected, contradicts the classical no-hair theorem \cite{Misner:1973prb}. The no-hair theorem states that, classically, the only data recoverable from the radiation state of the scalar field is the total mass $M$, arguing that the details of the collapse and its dynamics are irrelevant to obtain the Hawking radiation. This is not the case if the matter that forms the black hole is considered to be excitations of a quantised scalar field.

\section{Conclusion} \label{sec:conclusions}
This paper analyzed how non-vacuum initial quantum states influence Hawking radiation in black hole spacetimes. We classified excitations of the ``in'' vacuum into three categories: unitary operations within the black hole interior (``int'') algebra, non-unitary operations within this same algebra, and operations within the exterior null-infinity (``out'') algebra. Our results demonstrate that only unitary ``int'' operations leave no trace on the black hole radiation, while non-unitary ``int'' operations and both unitary and non-unitary ``out'' operations imprint (in principle) distinguishable perturbations onto outgoing radiation.
We invoked the AQFT localisation scheme to interpret ``int'' and ``out'' excitations as local operations in the causal-past of the black hole interior and exterior respectively.
The Knight-Licht strict localisation played an important role by showing that non-unitarily prepared states cannot be strictly localized, thereby clarifying which infalling quantum states are recoverable.
Along these lines, we considered the cases of single particle infalling states, and showed that those with different energies are not local-unitarily equivalent, and therefore distinguishable. We moreover compared infalling matter with a fixed mass $M$, in the cases of single particle vs two particles with $M/2$ each.
In general, we found that while complete information about infalling states is not recoverable from Hawking radiation, it can be retrieved up to local int-region unitaries.
Our analysis showed that stimulated emission significantly improves upon Hawking's original computations by demonstrating partial information recovery. 
These findings address aspects of the black hole information paradox by characterising the mechanisms for partial information recovery grounded in a rigorous algebraic quantum field theoretic formalism. However, complete information recovery, as well as the pure-to-mixed state evolution problem, have yet to be addressed building upon these results.
\vspace{4cm}

\appendix

\onecolumngrid

\section{Proof of Eq.\ (\ref{a_omega_neq_a_omega_prime})}\label{Proof_appendix}

Here, we prove that $N_{\omega} \hat{a}_{\omega}^{\text{int } \dagger} \ket{ {\text{in}}}\neq \left(U_{\text{int}} \otimes \mathbb{I}_{\text{out}} \right) N_{\omega '}\hat{a}_{\omega '}^{\text{int } \dagger} \ket{ {\text{in}}}$ for any $\omega \neq \omega '$, $N_{\omega},N_{\omega'}$ normalization factors, and any unitary that is local in the interior region $U_{\text{int}}$. Our proof is by contradiction. Thus, assume that there exist $\omega_1\neq \omega_2$ and a unitary $U_{\text{int}}$ such that  $ N_{\omega_1} \hat{a}_{\omega_1}^{\text{int } \dagger} \ket{ {\text{in}}} = \left(U_{\text{int}} \otimes \mathbb{I}_{\text{out}} \right) N_{\omega_2}\hat{a}_{\omega_2}^{\text{int } \dagger} \ket{ {\text{in}}}$. 
In order to show the contradiction, we first write the incoming vacuum state $\ket{ {\text{in}}}$ in terms of the int and out vacuum states and int and out creation operators. We know from the previous sections \cite{Fabbri} that 
\begin{gather}
    \ket{ {\text{in}}} = \bigotimes_{\omega} \sqrt{1-e^{-2 M \omega}} \sum_{n_{\omega}=0}^{\infty} e^{-M \omega n_{\omega}} \ket{n_\omega}_{\text{int}} \otimes \ket{n_\omega}_{\text{out}}.
\end{gather}

For convenience, we define $r_\omega \equiv e^{-2 M \omega}$. Notice that $r_\omega \in (0,1]$. It is straightforward to check that the well-normalised states we need are 
\be
N_{\omega_j} \hat{a}^{\text{int } \dagger}_{\omega_j}\ket{ {\text{in}}} && =
    \left( \bigotimes_{\omega \neq \omega_j} \sqrt{1-r_\omega} \sum_{n_{\omega}=0}^{\infty} r_{\omega}^{\frac{n_{\omega}}{2}} \ket{n_\omega}_{\text{int}} \otimes \ket{n_\omega}_{\text{out}} \right) \nonumber \\  && 
     \hspace{1cm } \otimes  \left(\left(1-r_{\omega_j}\right) \sum_{n_{\omega_j}=0}^{\infty} r_{\omega_j}^{\frac{n_{\omega_j}}{2}} \sqrt{n_{\omega_j}+1}\ket{n_{\omega_j}+1}_{\text{int}} \otimes \ket{n_{\omega_j}}_{\text{out}} \right)  \label{eq:singlexc}
\ee

Any local int unitary can be described without loss of generality as
\begin{gather}
U_{\text{int}} \otimes \mathbb{I}_{\text{out}}=\prod_\omega \sum_{n_\omega, m_\omega=0}^{\infty}   w\left(\{n_\omega\}_\omega, \{m_\omega\}_\omega\right) \bigotimes_{\omega} \ketbra{m_\omega}{n_\omega}_{\text{int}} \otimes  \mathbb{I}_{\text{out}}, \label{eq:unint}
\end{gather}
where $ w\left(\{n_\omega\}_\omega, \{m_\omega\}_\omega\right)$ are the coefficients of the local unitary operator, and $\{\bigotimes_{\omega} \ket{m_\omega}_{\text{int}}\}_{m_\omega=0, \omega}^{\infty}$ is the basis of the int space as the tensor product of Fock basis for each $\omega$.  The important properties are that the unitary is local and that the absolute value of the coefficients satisfies $ \left | w\left(\{n_\omega\}_\omega, \{m_\omega\}_\omega\right) \right | \leq 1$.

We are ready to obtain our contradiction. If the local unitary connects both states, we would have 
\begin{gather}
   N_{\omega_1} N_{\omega_2} \bra{ {\text{in}}} \hat{a}_{\omega_1}^{\text{int}} \left(U_{\text{int}} \otimes \mathbb{I}_{\text{out}} \right) \hat{a}_{\omega_2}^{\text{int } \dagger} \ket{ {\text{in}}} =1 
\end{gather}

To get the contradiction, let us expand the left-hand side and simplify: 
\begin{gather}
    N_{\omega_1} N_{\omega_2} \bra{ {\text{in}}} \hat{a}_{\omega_1}^{\text{int}} \left(U_{\text{int}} \otimes \mathbb{I}_{\text{out}} \right) \hat{a}_{\omega_2}^{\text{int } \dagger} \ket{ {\text{in}}} =  \nonumber \\ =  \left(\bigotimes_{\omega \neq \omega_1} \sqrt{1-r_\omega} \sum_{l_{\omega}=0}^{\infty} r_{\omega}^{\frac{l_{\omega}}{2}} \bra{l_\omega}_{\text{int}} \otimes \bra{l_\omega}_{\text{out}} \right)\otimes  \left(\left(1-r_{\omega_1}\right) \sum_{l_{\omega_1}=0}^{\infty} r_{\omega_1}^{\frac{l_{\omega_1}}{2}} \sqrt{l_{\omega_1}+1}\bra{l_{\omega_1}+1}_{\text{int}} \otimes \bra{l_{\omega_1}}_{\text{out}} \right)  \cdot \nonumber \\ \cdot \left(\prod_\omega \sum_{n_\omega,m_\omega=0}^{\infty}   w\left(\{n_\omega\}_\omega, \{m_\omega\}_\omega\right) \bigotimes_{\omega} \ketbra{m_\omega}{n_\omega}_{\text{int}} \otimes  \mathbb{I}_{\text{out}} \right) \cdot \nonumber \\ \cdot \left(\bigotimes_{\omega \neq \omega_2} \sqrt{1-r_\omega} \sum_{k_{\omega}=0}^{\infty} r_{\omega}^{\frac{k_{\omega}}{2}} \ket{k_\omega}_{\text{int}} \otimes \ket{k_\omega}_{\text{out}} \right)\otimes  \left(\left(1-r_{\omega_2}\right) \sum_{k_{\omega_2}=0}^{\infty} r_{\omega_2}^{\frac{k_{\omega_2}}{2}} \sqrt{k_{\omega_2}+1}\ket{k_{\omega_2}+1}_{\text{int}} \otimes \ket{k_{\omega_2}}_{\text{out}} \right)  \\ =  \prod_{\omega\neq \omega_1, \omega_2}  \left(1-r_\omega\right) \sum_{k_\omega=0}^{\infty} r_{\omega}^{k_\omega} \cdot \left(1-r_{\omega_1}\right)^{\frac{3}{2}}\cdot \left(1-r_{\omega_2}\right)^{\frac{3}{2}} \sum_{k_{\omega_1}, k_{\omega_2}=0}^{\infty} r_{\omega_1}^{k_{\omega_1}} r_{\omega_2}^{k_{\omega_2}} \sqrt{k_{\omega_1}+1} \sqrt{k_{\omega_2}+1} \cdot \nonumber \\  \cdot  w\left(\{k_\omega\}_{\omega\neq \omega_2}, k_{\omega_2} +1, \{k_\omega\}_{\omega\neq \omega_1}, k_{\omega_1}+1\right) =1
\end{gather}

Notice all the terms in the multisum are positive except for the unitary coefficients. Moreover, this is the only term that mixes coefficients of different $\omega$. So, now, since we have that the above sum is equal to $1$, by applying the inequality $ \left | w\left(\{n_\omega\}_\omega, \{m_\omega\}_\omega\right) \right | \leq 1$. We obtain an upper bound for $1$, as:
\begin{gather}
   1 \! \leq \! \! \left(\prod_{\omega\neq \omega_1, \omega_2}  \hspace{-3mm} \left(1-r_\omega\right) \sum_{k_\omega =0}^{\infty} r_{\omega}^{k_\omega} \right) \! \! \cdot \! \!  \left(\!\! \left(1-r_{\omega_1}\right)^{\frac{3}{2}} \! \! \! \sum_{k_{\omega_1}=0}^{\infty} \! \! r_{\omega_1}^{k_{\omega_1}} \! \sqrt{k_{\omega_1}+1}\! \right)\! \! \! \cdot \! \! \! \left( \! \! \left(1-r_{\omega_2}\right)^{\frac{3}{2}}  \! \! \sum_{k_{\omega_2}=0}^{\infty}  \! \!\!r_{\omega_2}^{k_{\omega_2}}  \! \sqrt{k_{\omega_2}+1} \right)
\end{gather}
Using the well-known $\sum_{k_\omega =0}^{\infty} r_{\omega}^{k_\omega}=\left(1-r_\omega\right)^{-1}$ and $\sum_{k_{\omega_j}=0}^{\infty}  r_{\omega_j}^{k_{\omega_2}}  \sqrt{k_{\omega_2}+1} = \frac{\text{Li}_{-\frac{1}{2}}(r_{\omega_j}) }{r_{\omega_j}}$; we obtain:
\begin{gather}
    1\leq  \left(1-r_{\omega_1}\right)^{\frac{3}{2}} \frac{\text{Li}_{-\frac{1}{2}}(r_{\omega_1}) }{r_{\omega_1}} \left(1-r_{\omega_2}\right)^{\frac{3}{2}} \frac{\text{Li}_{-\frac{1}{2}}(r_{\omega_2}) }{r_{\omega_2}}
\end{gather}
where $\text{Li}_{-\frac{1}{2}}(x)$ is the polylogarithm, or Jonquière's function, of order $-\frac{1}{2}$. Here we have arrived at our contradiction. It can be checked that the function $ \left(1-r\right)^{\frac{3}{2}} \frac{\text{Li}_{-\frac{1}{2}}(r) }{r} $ is positive and strictly smaller than $1$ in the range of $r\in (0,1]$. Therefore, we have a contradiction since we obtained that a function strictly smaller than $1$ is larger or equal to $1$. Thus, the two excited states we considered are not connected by a local int unitary. Therefore their two radiation states are different.

\section{Proof of Eq.\ (\ref{a_omega_neq_a_omega_prime_2})}\label{Proof_appendix_2}

Here, we prove Eq.\ (\ref{a_omega_neq_a_omega_prime_2}) by using a method analogous to the one in Appendix \ref{Proof_appendix}. We start by evaluating the states
\be
    N_M \hat{a}^{\text{int } \dagger}_{M} \ket{ {\text{in}}} && = \left(\bigotimes_{\omega \neq M} \sqrt{1-r_\omega} \sum_{n_{\omega}=0}^{\infty} r_{\omega}^{\frac{n_{\omega}}{2}} \ket{n_\omega}_{\text{int}} \otimes \ket{n_\omega}_{\text{out}} \right) \nonumber \\ 
    && \hspace{1.2cm} \otimes  \left(\left(1-r_{M}\right) \sum_{n_{M}=0}^{\infty} r_{M}^{\frac{n_{M}}{2}} \sqrt{n_{M}+1}\ket{n_{M}+1}_{\text{int}} \otimes \ket{n_{M}}_{\text{out}} \right)  \\ N_{\frac{M}{2} \frac{M}{2}} \left(\hat{a}^{\text{int } \dagger}_{\frac{M}{2}} \right)^2 \ket{ {\text{in}}} && = \left(\bigotimes_{\omega \neq \frac{M}{2}} \sqrt{1-r_\omega} \sum_{n_{\omega}=0}^{\infty} r_{\omega}^{\frac{n_{\omega}}{2}} \ket{n_\omega}_{\text{int}} \otimes \ket{n_\omega}_{\text{out}} \right) \nonumber \\ && \hspace {1.2 cm} \otimes  \left( \frac{\left(1-r_{\frac{M}{2}}\right)^{\frac{3}{2}}}{\sqrt{2}} \sum_{k=0}^{\infty} r_{\frac{M}{2}}^{\frac{k}{2}} \sqrt{k+1}\sqrt{k+2}\ket{k+2}_{\text{int}} \otimes \ket{k}_{\text{out}} \right)
\ee
where we have used Eq. \ref{eq:singlexc}. We now proceed in the same fashion as in the previous subsection. Thus, we follow the same logical reasoning with a contradiction argument. We use Eq. \ref{eq:unint} to calculate:
\be
    && N_{\frac{M}{2} \frac{M}{2}}  N_M  \bra{ {\text{in}}} \hat{a}^{\text{int }}_{M} \left(U_{\text{int}} \otimes \mathbb{I}_{\text{out}}\right) \left(\hat{a}^{\text{int } \dagger}_{\frac{M}{2}} \right)^2 \ket{ {\text{in}}}
    \nonumber \\ 
     &&= \left(\bigotimes_{\omega \neq M} \sqrt{1-r_\omega} \sum_{l_{\omega}=0}^{\infty} r_{\omega}^{\frac{l_{\omega}}{2}} \bra{l_\omega}_{\text{int}} \otimes \bra{l_\omega}_{\text{out}} \right)\otimes  \left(\left(1-r_{M}\right) \sum_{l_{M}=0}^{\infty} r_{M}^{\frac{l_{M}}{2}} \sqrt{l_{M}+1}\bra{l_{M}+1}_{\text{int}} \otimes \bra{l_{M}}_{\text{out}} \right) \cdot \nonumber \\  
     && \hspace{3cm} \cdot \left(\prod_\omega \sum_{n_\omega, m_\omega=0}^{\infty}   w\left(\{n_\omega\}_\omega, \{m_\omega\}_\omega\right) \bigotimes_{\omega} \ketbra{m_\omega}{n_\omega}_{\text{int}} \otimes  \mathbb{I}_{\text{out}} \right) \cdot \nonumber \\ 
     && \cdot \left(\bigotimes_{\omega \neq \frac{M}{2}} \sqrt{1-r_\omega} \sum_{k_{\omega}=0}^{\infty} r_{\omega}^{\frac{k_{\omega}}{2}} \ket{k_\omega}_{\text{int}} \otimes \ket{k_\omega}_{\text{out}} \right)\otimes  \left( \frac{\left(1-r_{\frac{M}{2}}\right)^{\frac{3}{2}}}{\sqrt{2}} \sum_{k=0}^{\infty} r_{\frac{M}{2}}^{\frac{k}{2}} \sqrt{k+1}\sqrt{k+2}\ket{k+2}_{\text{int}} \otimes \ket{k}_{\text{out}} \right) \nonumber \\ 
     && =  \prod_{\omega\neq M, \frac{M}{2}}  \left(1-r_\omega\right) \sum_{k_\omega=0}^{\infty} r_{\omega}^{k_\omega} \cdot \left(1-r_{M}\right)^{\frac{3}{2}}\cdot \frac{\left(1-r_{\frac{M}{2}}\right)^{2}}{\sqrt{2}} \sum_{k_{M}, k=0}^{\infty} r_{M}^{k_{M}} r_{\frac{M}{2}}^{k} \sqrt{k_{M}+1} \sqrt{k+1} \sqrt{k+2} \cdot \nonumber \\  
     &&\hspace{5cm}  \cdot  w\left(\{k_\omega\}_{\omega\neq \frac{M}{2}}, k +2, \{k_\omega\}_{\omega\neq M}, k_{M}+1\right) \\ 
     && =1
\ee
With the equality holding if the local int unitary that connects the two states exists. Now we can upper bound such expression using the fact that the unitary coefficients are individually bounded by 1. Thus, using $\sum_{k_\omega =0}^{\infty} r_{\omega}^{k_\omega}=\left(1-r_\omega\right)^{-1}$ and $\sum_{k_{\omega_j}=0}^{\infty}  r_{\omega_j}^{k_{\omega_2}}  \sqrt{k_{\omega_2}+1} = \frac{\text{Li}_{-\frac{1}{2}}(r_{\omega_j}) }{r_{\omega_j}}$ we obtain
\begin{gather}
    1 \leq \left(1-r_{M}\right)^{\frac{3}{2}} \frac{\text{Li}_{-\frac{1}{2}}(r_{M}) }{r_{M}}  \frac{\left(1-r_{\frac{M}{2}}\right)^{2}}{\sqrt{2}} \sum_{ k=0}^{\infty} r_{\frac{M}{2}}^{k} \sqrt{k+1} \sqrt{k+2} \leq \nonumber \\ \leq \left(1-r_{M}\right)^{\frac{3}{2}} \frac{\text{Li}_{-\frac{1}{2}}(r_{M}) }{r_{M}}  \left(1-r_{\frac{M}{2}}\right)^{2} \sum_{ k=0}^{\infty} r_{\frac{M}{2}}^{k} \left(\frac{3}{4}k +1\right)= \left(1-r_{M}\right)^{\frac{3}{2}} \frac{\text{Li}_{-\frac{1}{2}}(r_{M}) }{r_{M}}  \left(1-\frac{r}{4}\right)
\end{gather}
And again, we see the function in the right hand side is strictly smaller than 1 in the whole range of $r\in(0,1]$.

\section{The Coherent State}
\label{app:The_Coherent_State}

As an example of a unitary excitation of the ``in'' vacuum,  consider coherent states, defined by
\begin{equation}
O_\text{in} (\hat{a}^\text{in}_\omega, \hat{a}^{\text{in} \dagger}_\omega) = \exp ( \int_0^{+\infty} d\omega \psi_\omega \hat{a}^{\text{in} \dagger}_\omega - \text{h.c.}).
\end{equation}
In this case, the operator $\hat{O}$ is unitary. Hence, the condition for the state $| \psi \rangle$  to be indistinguishable from the ``in'' vacuum $| \text{in} \rangle$ in the ``out'' region is that $\hat{O}$ must consist entirely of ``int'' annihilation and creation operators. Here, we investigate how this condition can be expressed in terms of the function $\psi_\omega$.

By following the procedure outlined in Sec.\ \ref{sec:Non-vacuum}, the ``in'' coherent state $| \psi \rangle$ can be expressed in terms of ``out'' and ``int'' particles as in Eq.\ (\ref{psi_O_out_int}), with
\begin{align}
& O_\text{out,int} \left( \hat{a}^\text{out}_\omega, \hat{a}^{\text{out} \dagger}_\omega,\hat{a}^\text{int}_\omega, \hat{a}^{\text{int} \dagger}_\omega \right) = \exp \left( \hat{O}_\text{out} - \text{h.c.} \right)  \otimes \exp \left(  \hat{O}_\text{int}  - \text{h.c.} \right), \nonumber \\
& \hat{O}_\text{out} = \int_0^{+\infty} d\omega \int_0^{+\infty} d\omega' \psi_\omega \left[ \left(  u^\text{out}_{\omega'} , u^\text{in}_\omega \right)  \hat{a}^\text{out}_{\omega'} + \left(  u^{\text{out} *}_{\omega'} , u^\text{in}_\omega \right)  \hat{a}^{\text{out}  \dagger}_{\omega'}  \right], \nonumber \\
& \hat{O}_\text{int} = \int_0^{+\infty} d\omega \int_0^{+\infty} d\omega' \psi_\omega \left[ \left(  u^\text{int}_{\omega'} , u^\text{in}_\omega \right) \hat{a}^\text{int}_{\omega'}  + \left(  u^{\text{int}*}_{\omega'} , u^\text{in}_\omega \right) \hat{a}^{\text{int} \dagger}_{\omega'}  \right].
\end{align}
By tracing out the ``int'' degrees of freedom and using the cyclic property of the trace along with the unitarity of $\exp (  \hat{O}_\text{int}  - \text{h.c.} )$, we obtain 
\begin{equation}
\text{Tr}_\text{int}  | \psi \rangle  \langle \psi | = \exp \left( \hat{O}_\text{out} - \text{h.c.} \right) \hat{\rho}_\text{H} \exp \left( - \hat{O}_\text{out} - \text{h.c.} \right).
\end{equation}

The condition for the field perturbation to not affect the Hawking radiation is $\hat{O}_\text{out} = 0$, which is equivalent to
\begin{align}\label{condition_unaffected_Hawking}
& \left(  u^\text{out}_{\omega'} , \int_0^{+\infty} d\omega \psi_\omega u^\text{in}_\omega \right) = 0 , \\ &
\hspace{2cm} \text{and} \nonumber \\ & 
\left(  u^{\text{out} *}_{\omega'} , \int_0^{+\infty} d\omega \psi_\omega u^\text{in}_\omega \right) = 0 , ~~ \text{for any } \omega'.
\end{align}
To evaluate the Klein-Gordon inner product in Eq.\ (\ref{condition_unaffected_Hawking}), we consider the surface $\mathcal{I}^+ \cup \mathcal{H}^+$. Since $u^\text{out}_{\omega}$ and $u^\text{int}_{\omega}$ form a complete basis for modes supported on $\mathcal{I}^+ $ and $ \mathcal{H}^+$, respectively, Eq.\ (\ref{condition_unaffected_Hawking}) is equivalent to requiring that the mode $\int_0^{+\infty} d\omega \psi_\omega u^\text{in}_\omega$ is entirely supported on $ \mathcal{H}^+$ when restricted to $\mathcal{I}^+ \cup \mathcal{H}^+$, meaning that
\begin{equation}\label{condition_unaffected_Hawking_2}
\left. \int_0^{+\infty} d\omega \psi_\omega u^\text{in}_\omega \right|_{\mathcal{I}^+} (u_\text{out}) = 0, \qquad \text{for any } u_\text{out}.
\end{equation}
When Eq.\ (\ref{condition_unaffected_Hawking_2}) is satisfied, the excitation $| \psi \rangle$ does not perturb the Hawking radiation. This condition is both necessary and sufficient.

To explicitly express Eq.\ (\ref{condition_unaffected_Hawking_2}) for the specific case of the Vaidya spacetime, we use Eq.\ (\ref{u_in_S}), which describes $u^\text{in}_\omega$ in the Schwarzschild region $\mathcal{S}$. In the limit $v \to \infty$, the mode behaves as $\left. u^\text{in}_\omega \right|_\mathcal{S}(v) \to - e^{-i \omega u_\text{in}(u_\text{out},v_0)}/4 \pi \sqrt{\omega} r$. Substituting this into Eq.\ (\ref{condition_unaffected_Hawking_2}), we obtain the explicit condition
\begin{equation}\label{condition_unaffected_Hawking_3}
\int_0^{+\infty} d\omega   \psi_\omega  \frac{e^{-i \omega u_\text{in}(u_\text{out},v_0)}}{4 \pi \sqrt{\omega} r} = 0, \qquad \text{for any } u_\text{out}.
\end{equation}
This therefore is the necessary and sufficient condition for a coherent state excitation of the ``in'' vacuum to be undistinguishable from the non-excited ``in'' vacuum state, through the probation of the outgoing Hawking radiation, in the case of a Vaidya spacetime. 

\bibliography{BH}

\begin{thebibliography}{10}

\bibitem{Hawking74}
S.~W. Hawking.
\newblock Particle creation by black holes.
\newblock {\em Commun. Math. Phys.}, 43:199, 1975.

\bibitem{Parker:1975jm}
L.~Parker.
\newblock {Probability Distribution of Particles Created by a Black Hole}.
\newblock {\em Phys. Rev. D}, 12:1519--1525, 1975.

\bibitem{BirrelDavies}
N.~D. Birrell and P.~C.~W. Davies.
\newblock {\em Quantum Fields in Curved Space}.
\newblock Cambridge University Press, 1982.

\bibitem{Fabbri}
Alessandro Fabbri and Jose Navarro-Salas.
\newblock {\em Modeling Black Hole Evaporation}.
\newblock Imperial College Press, 2005.

\bibitem{Hawking76}
S.~W. Hawking.
\newblock Black holes and thermodynamics.
\newblock {\em Phys. Rev. D}, 13:191, 1976.

\bibitem{HawkingInfo}
S.~W. Hawking.
\newblock Breakdown of predictability in gravitational collapse.
\newblock {\em Phys. Rev. D}, 14:2460, 1976.

\bibitem{Adler:2001vs}
R.~J. Adler, P.~Chen, and D.~I. Santiago.
\newblock The generalized uncertainty principle and black hole remnants.
\newblock {\em Gen. Rel. Grav.}, 33:2101--2108, 2001.

\bibitem{Stable}
P.~Chen, Y.~C. Ong, and D.~h.~Yeom.
\newblock Black hole remnants and the information loss paradox.
\newblock {\em Phys. Rept.}, 603:1--45, 2015.

\bibitem{Perez}
T.~De Lorenzo and A.~Perez.
\newblock Improved black hole fireworks: Asymmetric black-hole-to-white-hole tunneling scenario.
\newblock {\em Phys.\ Rev.\ D}, 93(12):124018, 2016.

\bibitem{Rovelli2}
H.~M. Haggard and C.~Rovelli.
\newblock Black hole fireworks: quantum-gravity effects outside the horizon spark black to white hole tunneling.
\newblock {\em Phys. Rev. D}, 92:104020, 2015.

\bibitem{Susskind1}
L.~Susskind and J.~Lindesay.
\newblock {\em An introduction to black holes, information and the string theory revolution: The holographic universe}.
\newblock World Scientific, 2005.

\bibitem{Susskind2}
L.~Susskind and L.~Thorlacius.
\newblock Gedanken experiments involving black holes.
\newblock {\em Phys.\ Rev.\ D}, 49:966, 1994.

\bibitem{Complementarity}
L.~Susskind, L.~Thorlacius, and J.~Uglum.
\newblock The stretched horizon and black hole complementarity.
\newblock {\em Phys.\ Rev.\ D}, 48:3743, 1993.

\bibitem{Maldacena}
G.~T. Horowitz and J.~M. Maldacena.
\newblock The black hole final state.
\newblock {\em JHEP}, 0402:008, 2004.

\bibitem{Lloyd}
S.~Lloyd and J.~Preskill.
\newblock Unitarity of black hole evaporation in final-state projection models.
\newblock {\em JHEP}, 1408:126, 2014.

\bibitem{Harlow}
D.~Harlow and P.~Hayden.
\newblock Quantum computation vs. firewalls.
\newblock {\em JHEP}, 06:085, 2013.

\bibitem{Mathur}
S.D.Mathur.
\newblock The information paradox: A pedagogical introduction.
\newblock {\em Class.Quant.Grav.}, 26:224001, 2009.

\bibitem{AMPS}
A.~Almheiri, D.~Marolf, J.~Polchinski, and J.~Sully.
\newblock Black holes: Complementarity or firewalls?
\newblock {\em JHEP}, 1302:062, 2013.

\bibitem{Braunstein}
K.~Zyczkowski S.~L.~Braunstein, S.~Pirandola.
\newblock Better late than never: Information retrieval from black holes.
\newblock {\em Phys. \ Rev. \ Lett.}, 110(10):101301, 2013.

\bibitem{Penington}
G.~Penington.
\newblock Entanglement wedge reconstruction and the information paradox.
\newblock {\em JHEP}, 09:002, 2020.

\bibitem{Almheiri:2019qdq}
A.~Almheiri, T.~Hartman, J.~Maldacena, E.~Shaghoulian, and A.~Tajdini.
\newblock Replica wormholes and the entropy of hawking radiation.
\newblock {\em JHEP}, 05:013, 2020.

\bibitem{Penington:2019kki}
G.~Penington, S.~H. Shenker, D.~Stanford, and Z.~Yang.
\newblock Replica wormholes and the black hole interior.
\newblock {\em JHEP}, 03:205, 2022.

\bibitem{Akil1}
A.~Akil, O.~Dahlsten, and L.~Modesto.
\newblock Conditional entanglement transfer via black holes: restoring predictability.
\newblock {\em New J. Phys.}, 23(11):113011, 2021.

\bibitem{BOOK}
A.~Akil and C.~Bambi(Eds).
\newblock {\em {The Black Hole Information Paradox, A Fifty-years Journey}}.
\newblock Springer, Singapore, 2025.

\bibitem{Press:1972zz}
William~H. Press and Saul~A. Teukolsky.
\newblock {Floating Orbits, Superradiant Scattering and the Black-hole Bomb}.
\newblock {\em Nature}, 238:211--212, 1972.

\bibitem{Starobinsky:1973aij}
A.~A. Starobinsky.
\newblock {Amplification of waves reflected from a rotating ''black hole''.}
\newblock {\em Sov. Phys. JETP}, 37(1):28--32, 1973.

\bibitem{Wald:1976ka}
Robert~M. Wald.
\newblock {Stimulated Emission Effects in Particle Creation Near Black Holes}.
\newblock {\em Phys. Rev. D}, 13:3176--3182, 1976.

\bibitem{Bekenstein:1977mv}
J.~D. Bekenstein and A.~Meisels.
\newblock {Einstein a and B Coefficients for a Black Hole}.
\newblock {\em Phys. Rev. D}, 15:2775--2781, 1977.

\bibitem{Panangaden:1977pc}
P.~Panangaden and Robert~M. Wald.
\newblock {Probability Distribution for Radiation from a Black Hole in the Presence of Incoming Radiation}.
\newblock {\em Phys. Rev. D}, 16:929--932, 1977.

\bibitem{Sorkin:1986zj}
Rafael Sorkin.
\newblock {A Simple Derivation of Stimulated Emission by Black Holes}.
\newblock {\em Class. Quant. Grav.}, 4:L149, 1987.

\bibitem{Audretsch:1994ga}
Jurgen Audretsch and Rainer Muller.
\newblock {Localized discussion of stimulated processes for Rindler observers and accelerated detectors}.
\newblock {\em Phys. Rev. D}, 49:4056--4065, 1994.

\bibitem{Brandenberger:1992sr}
Robert~H. Brandenberger, Viatcheslav~F. Mukhanov, and T.~Prokopec.
\newblock {Entropy of a classical stochastic field and cosmological perturbations}.
\newblock {\em Phys. Rev. Lett.}, 69:3606--3609, 1992.

\bibitem{Keski-Vakkuri:1993hsn}
Esko Keski-Vakkuri.
\newblock {On coarse grained entropy and stimulated emission in curved space-time}.
\newblock {\em Phys. Rev. D}, 49:2122--2125, 1994.

\bibitem{Vendrell:1996am}
F.~Vendrell.
\newblock {Stimulated emission of particles by (1+1)-dimensional black holes}.
\newblock {\em Helv. Phys. Acta}, 70:637--669, 1997.

\bibitem{Lochan:2015oba}
Kinjalk Lochan and T.~Padmanabhan.
\newblock {Extracting information about the initial state from the black hole radiation}.
\newblock {\em Phys. Rev. Lett.}, 116(5):051301, 2016.

\bibitem{Lochan:2016nbs}
Kinjalk Lochan, Sumanta Chakraborty, and T.~Padmanabhan.
\newblock {Information retrieval from black holes}.
\newblock {\em Phys. Rev. D}, 94(4):044056, 2016.

\bibitem{Bradler:2013gqa}
Kamil Br\'adler and Christoph Adami.
\newblock {The capacity of black holes to transmit quantum information}.
\newblock {\em JHEP}, 05:095, 2014.

\bibitem{Jacobson:2003vx}
Ted Jacobson.
\newblock {Introduction to quantum fields in curved space-time and the Hawking effect}.
\newblock In {\em {School on Quantum Gravity}}, pages 39--89, 8 2003.

\bibitem{haag1992local}
R.~Haag.
\newblock {\em Local Quantum Physics: Fields, Particles, Algebras}.
\newblock R.Balian, W.Beiglbock, H.Grosse. Springer-Verlag, 1992.

\bibitem{Schrodinger1936}
E.~Schrödinger.
\newblock Probability relations between separated systems.
\newblock {\em Mathematical Proceedings of the Cambridge Philosophical Society}, 32(3):446–452, 1936.

\bibitem{HUGHSTON199314}
Lane~P. Hughston, Richard Jozsa, and William~K. Wootters.
\newblock A complete classification of quantum ensembles having a given density matrix.
\newblock {\em Physics Letters A}, 183(1):14--18, 1993.

\bibitem{10.1063/1.1703731}
James~M. Knight.
\newblock {Strict Localization in Quantum Field Theory}.
\newblock {\em Journal of Mathematical Physics}, 2(4):459--471, 12 1961.

\bibitem{10.1063/1.1703925}
A.~L. Licht.
\newblock {Strict Localization}.
\newblock {\em Journal of Mathematical Physics}, 4(11):1443--1447, 1963.

\bibitem{Ric}
Riccardo Falcone and Claudio Conti.
\newblock {Localization in quantum field theory}.
\newblock {\em Rev. Phys.}, 12:100095, 2024.

\bibitem{Misner:1973prb}
Charles~W. Misner, K.~S. Thorne, and J.~A. Wheeler.
\newblock {\em {Gravitation}}.
\newblock W. H. Freeman, San Francisco, 1973.

\end{thebibliography}

\end{document}